\documentclass[sigconf,edbt,dvipsnames]{acmart-edbt2018}

\renewcommand\footnotetextcopyrightpermission[1]{} 
\usepackage{booktabs} 

\usepackage{xcolor}

\usepackage{amssymb}
\usepackage{amsmath}
\usepackage{graphicx}
\graphicspath{ {Figures/} }
\usepackage{balance}
\usepackage{wrapfig}

\usepackage{multirow}
\usepackage{pifont}
\usepackage{marvosym,etoolbox}

\usepackage{subcaption}
\usepackage{url}
\usepackage{float}
\usepackage{verbatim}
\usepackage{adjustbox}

\newtheorem{observation}{Observation}

\setcopyright{rightsretained}
\acmDOI{}
\acmISBN{978-3-89318-078-3}
\acmConference[EDBT 2018]{21st International Conference on Extending Database Technology (EDBT)}{March 26-29, 2018}{Vienna, Austria} 
\acmYear{2018}
\begin{document}

\title[MLBench: Benchmarking Machine Learning Clouds]{MLBench: How Good Are Machine Learning Clouds for Binary Classification Tasks on Structured Data?}
\subtitle{[Experiments and Analysis]}

\let\OLDthebibliography\thebibliography
\renewcommand\thebibliography[1]{
  \OLDthebibliography{#1}
  \setlength{\parskip}{0pt}
  \setlength{\itemsep}{0pt plus 0.3ex}
}


\author{Yu Liu}
\affiliation{%
  \institution{Department of Computer Science\\ETH Zurich}
}
\email{yu.liu@inf.ethz.ch}

\author{Hantian Zhang}
\affiliation{%
  \institution{Department of Computer Science\\ETH Zurich}
}
\email{hantian.zhang@inf.ethz.ch}

\author{Luyuan Zeng}
\affiliation{%
  \institution{Department of Computer Science\\ETH Zurich}
}
\email{zengl@student.ethz.ch}

\author{Wentao Wu}
\affiliation{%
  \institution{Microsoft Research, Redmond}
}
\email{wentao.wu@microsoft.com}

\author{Ce Zhang}
\affiliation{%
  \institution{Department of Computer Science\\ETH Zurich}
}
\email{ce.zhang@inf.ethz.ch}

\begin{abstract}

We conduct an empirical study of machine learning functionalities 
provided by major cloud service providers, which we call {\em machine learning clouds}.
Machine learning clouds hold the promise of hiding all the sophistication of running large-scale machine learning: Instead of specifying {\em how} to run a machine learning task, users only specify {\em what} machine learning task to run and the cloud figures out the rest.
Raising the level of abstraction, however, rarely comes free --- a performance penalty is possible.
{\em How good, then, are current
machine learning clouds on real-world machine learning workloads?}

We study this question with a focus on binary classification problems.
We present \texttt{mlbench}, a novel benchmark constructed by harvesting datasets from Kaggle competitions.
We then compare the performance of the top winning code available from Kaggle with that of running  machine learning clouds from both Azure and Amazon
on \texttt{mlbench}.
Our comparative study reveals the strength and weakness of existing machine learning clouds and points out potential future directions for improvement.

\end{abstract}

\maketitle

\vspace{-1.0em}
\section{Introduction}

In spite of the recent advancement of machine learning research, modern machine learning systems
are still far from easy to use, at least from the perspective of business users or even
scientists without a computer science background~\cite{zhang2017over}.
Recently, there is a trend toward pushing machine learning onto the cloud as a ``service,'' a.k.a. {\em machine learning clouds}.
By putting a set of machine learning primitives on the cloud, these services significantly
raise the level of abstraction for machine learning.
For example, with Amazon Machine Learning, users only need to upload the dataset and specify the type of task
(classification or regression).
The cloud will then automatically train machine learning models without any user intervention.

\vspace{0.5em}
\noindent
From a data management perspective, the emergence of machine learning clouds represents an attempt toward {\em declarative machine learning}. Instead of relying on users to specify {\em how} a machine learning task should be
configured, tuned, and executed, machine learning clouds manage all these physical decisions and allow users to focus on the logical side: {\em what} tasks they want to perform with machine learning.

Raising the level of abstractions and
building a system to automatically manage all
physical decisions, however, rarely comes free.
In the context of a data management system, a sophisticated query optimizer is responsible for generating good physical execution plans.
Despite the intensive and extensive research and engineering effort that has been put into building capable query optimizers in the past four decades, query optimizers still often make mistakes that lead to disastrous performance.

In the context of declarative machine learning, things become even more subtle. A bad choice of ``physical plan'' may result in not only suboptimal performance but also suboptimal quality (e.g., accuracy). In this paper, we investigate the usability of state-of-the-art machine learning clouds. Specifically, we ask the following question: 

\begin{description}
    \item {\em To what extent can existing declarative machine learning clouds support real-world machine learning tasks?}
\end{description}

More concretely, {\em what would users lose by resorting to declarative machine learning clouds instead of using 
a best-effort, non-declarative machine learning system?}
To answer this question, we conduct an empirical study with 
\texttt{mlbench}, a novel benchmark consisting of real-world datasets
{\em and} best-effort solutions
harvested from Kaggle competitions.
We use a novel methodology that allows us to separate measuring the performance of the machine learning clouds themselves from other external factors that may have significant impact
on quality, such as feature selection and hyper-parameter tuning.
Moreover, we use a novel performance metric that measures the strength and weakness of current machine learning clouds by comparing their relative performance with top-ranked solutions in Kaggle competitions.

\vspace{0.3em}
\noindent
{\bf Premier of Kaggle Competitions}. 
Kaggle is a popular platform 
hosting a range of machine learning 
competitions. 
Companies or scientists can host
their real-world applications
on Kaggle; Each user has access
to the training and testing datasets,
submits their solutions to Kaggle,
and gets a quality score on the
test set. Kaggle motivates users with prizes for top winning
entries. This ``crowdsourcing''
nature of Kaggle makes it a
representative sample of 
real-world machine learning workloads.

\vspace{0.3em}
\noindent
{\bf Summary of Technical Contributions}. 

\noindent
{\bf C1.} We present the \texttt{mlbench} benchmark. One prominent feature of \texttt{mlbench} is that each of its datasets comes with a best-effort baseline of both feature engineering and
selection of machine learning models.

\noindent
{\bf C2.} We propose a novel performance metric based on the notion of ``quality tolerance''
that measures the performance gap between a given machine learning cloud and top-ranked Kaggle competition performers.

\noindent
{\bf C3.} We evaluate the two most popular machine learning clouds, Azure Machine Learning Studio and Amazon Machine Learning, using \texttt{mlbench}. Our experimental result reveals interesting strengths and limitations of both clouds. Detailed analysis of the results further points out promising future directions
to improve both machine learning clouds.

\noindent
{\bf Overview}. 
The rest of the paper is organized as follows. We present our methodology in Section~\ref{sec:preliminaries} and the \texttt{mlbench} benchmark in Section~\ref{sec:benchmark}. We then present experimental settings and evaluation results in Section~\ref{sec:settings}-\ref{sec:results:feature-engineering}. We summarize related work in Section~\ref{sec:relatedwork} and conclude in Section~\ref{sec:conclusion}.

\section{Methodology}\label{sec:preliminaries}

Benchmarking systems fairly is not an easy task.
Three key aspects came to mind when
designing a benchmark for machine learning clouds:

\noindent
{\bf (1)} We need to measure not only the {\em performance} (speed) but also 
the {\em quality} (precision). The two are coupled, and their relative importance
changes with respect to the user's budget and tolerance for suboptimal quality. 

\noindent
{\bf (2)} The quality of an application depends on both {\em feature engineering} 
and the {\em machine learning model}. If these two factors are not decoupled, our result will be unfair to most machine learning clouds, as they usually do not provide an efficient mechanism for automatic feature engineering.

\noindent
{\bf (3)} To compare declarative machine learning clouds with the best effort of using non-declarative machine learning systems, we need to construct a {\em strong baseline} for the latter. If this baseline is not strong enough, our result may be overly optimistic regarding machine learning clouds.

Starting from these principles, we made a few basic decisions that we shall present next.

\subsection{Scope of Study} \label{sec:preliminaries:scope}

We restrict ourselves to {\em binary classification}, one of the most
popular machine learning tasks. As we will see, even with this
constrained scope, there is no simple, single answer
to the main question we aim to answer.

\subsection{Methodology} \label{sec:preliminaries:methodology}

We collect top winning code for all binary classification competitions on Kaggle. We then filter them to select 
a subset to include in \texttt{mlbench} with the following protocol. For the code that we are able to install and finish running within 24 hours, we further collect 
features extracted by the winning code.
The features are then used for training and testing models provided by both the machine learning cloud and the Kaggle winning solution. We also include datasets constructed
using raw features (see Section~\ref{sec:mlbench:overview}).

\paragraph*{Discussion} At first glance, our methodology is quite trivial.
Indeed, there is little novelty in the procedure itself, though the engineering effort involved is substantial. 
(It took us more than nine months to finish the experimental evaluation presented in Section~\ref{sec:results}.)
On second thought, one may wonder what the point is of spending so much effort.

To see the subtlety here, consider an alternative approach that is much easier to implement: take one well-known dataset (or several datasets)
such as those from the 
UCI Machine Learning Repository, run a standard feature selection algorithm, and compare the performance of machine learning clouds with that of standard machine learning libraries (e.g., Weka~\cite{hall2009weka}) on this dataset.
There are, however, a couple of caveats in this approach.
First, it is unclear how challenging the learning problem (associated with the dataset) is.
There may be {\em subjective} justification but no {\em objective} metric of the difficulty.
Second, it is questionable whether the models covered by standard libraries represent the state of the art.
Depending on the popularity and maturity of the libraries, coverage may vary dramatically.
Third, feature engineering and model selection are more of an art mastered only by human experts. If we ignore
both, our result might be overly optimistic or
overly pessimistic for machine learning clouds.

The intuition behind our methodology is simple: the top winning code of Kaggle competitions represents the arguably best effort among existing machine-learning solutions.
Of course, it is biased by the competitions published on Kaggle and the solutions provided by the participants.
Nonetheless, given the high impact of Kaggle competitions, we believe that using the winning code as a performance baseline significantly raises the bar compared with using standard libraries and therefore reduces the risk that we might be overly optimistic about the machine learning clouds.
Moreover, given the ``crowdsourcing'' nature of Kaggle, the baseline will keep up with the advancement of machine learning research and practice, perhaps at a much faster pace than standard libraries can.

\subsection{Quality Metric} \label{sec:preliminaries:metric}

Our methodology of adopting Kaggle winning code as a baseline raises the question of designing a reasonable quality 
metric. To measure the quality of a model deployed on machine learning clouds, we introduce the 
notion of ``quality tolerance'' (of a user).

\begin{definition}\label{def:quality-tolerance}
The \emph{quality tolerance} of a user is $\tau$ if s/he can be satisfied only by being ranked among the top $\tau$\%, assuming that s/he uses a model $M$ provided by the cloud to participate in a Kaggle competition.
\end{definition}

Of course, the ``user'' in Definition~\ref{def:quality-tolerance} is just hypothetical.
Essentially, quality tolerance measures the performance gap between the machine learning cloud and the top-ranked code of a Kaggle competition.
A lower quality tolerance suggests a more stringent user requirement and therefore a more capable machine learning cloud if it can meet that quality tolerance.

Based on the notion of quality tolerance, we are mainly interested in two performance metrics of a model $M$:
\begin{itemize}
    \item \emph{Capacity}, the minimum quality tolerance $\tau_{\min}$ that $M$ can meet for a given Kaggle competition $T$;
    \item \emph{Universality}, the number of Kaggle competitions that $M$ can achieve a quality tolerance of $\tau$.
\end{itemize}
Intuitively, capacity measures how high $M$ can be ranked in a Kaggle competition, whereas universality measures in how many Kaggle competitions $M$ can be ranked that high.

We use $c(M, T)$ and $u(M,\tau)$ to denote the capacity and $\tau$-universality of $M$.
Moreover, we use $\mathcal{K}(M, \tau)$ to denote the set of Kaggle competitions whose quality tolerance $\tau$ have been reached by $u(M, \tau)$: 
$$u(M, \tau)=|\mathcal{K}(M, \tau)|.$$

Similarly, if a machine learning cloud $\mathcal{M}$ provides $n$ models $\{M_1, ..., M_n\}$ ($n\geq 1$), we can define the capacity of $\mathcal{M}$ with respect to a Kaggle competition $T$ as
$$c(\mathcal{M}, T)=\min_{M_i\in\mathcal{M}}c(M_i, T),\quad 1\leq i\leq n,$$
and define the $\tau$-universality of $\mathcal{M}$ as
$$u(\mathcal{M},\tau)=|\bigcup\nolimits_{i=1}^{n}\mathcal{K}(M_i,\tau)|.$$
Clearly, the capacity of $\mathcal{M}$ over $T$ is the capacity of the best model that $\mathcal{M}$ provides for $T$, whereas the $\tau$-university of $\mathcal{M}$ is the number of Kaggle competitions in which $\mathcal{M}$ can meet quality tolerance $\tau$ (with the best model it can provide).

Finally, if there are $m$ Kaggle competitions $\mathcal{T}=\{T_1, ..., T_m\}$, we define the capacity of $\mathcal{M}$ over $\mathcal{T}$ as
$$c(\mathcal{M}, \mathcal{T})=\max_{T_j\in\mathcal{T}}c(\mathcal{M}, T_j),\quad 1\leq j\leq m.$$
It measures the {\em uniformly} best quality tolerance that $\mathcal{M}$ can meet for any of the competitions in $\mathcal{T}$.

In the rest of this paper, we will use the notation $c(M)$, $u(M)$, $c(\mathcal{M})$, and $u(\mathcal{M})$ whenever the corresponding quality tolerance and Kaggle competition(s) are clear from the context.

\begin{figure}[t]
\centering
    \includegraphics[width=0.48\columnwidth]{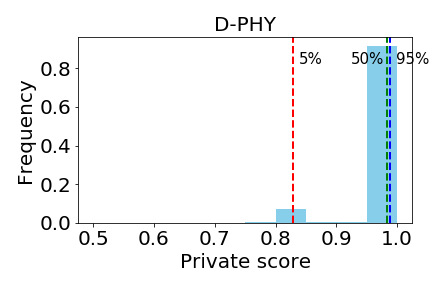}
    \includegraphics[width=0.48\columnwidth]{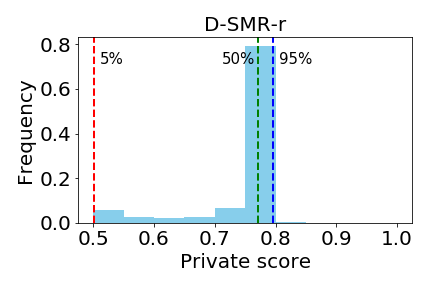}
\vskip -3ex
\caption{Histograms of the AUC scores on private leader board for two example datasets D-PHY and D-SMR-r.}
\label{fig:histogram}
\vskip -2ex
\end{figure}

\subsection{Limitations and Discussion}

Our motivation of using ranking as 
performance metric is to provide a normalized
score {\em across} all datasets. 
However, ranking itself does not tell the full story.
One caveat is that ranking measures the \emph{relative} performance and may be sensitive to the change in the underlying, \emph{absolute} metric, such as the ``area under curve'' (AUC) score that is commonly used by Kaggle competitions.
To illustrate this, Figure~\ref{fig:histogram}
presents the histograms (i.e., distributions) of the AUC scores in two Kaggle competitions (see Section~\ref{sec:benchmark:dataset-details} for the details of the competitions).
The red, green, and blue lines correspond
to the teams ranked at the top 95\%, 50\%, and 5\%.
The distance between the scores of 
top 50\% (green) and top 5\% (blue) shows the 
sensitivity --- for
D-PHY, ranking is quite sensitive to small changes
in AUC as most of the teams have similar scores.
Therefore, when benchmarking machine learning clouds,
it is important to look at {\em both} ranking
and absolute quality. In this paper,
our analysis will always base on both.

\section{The \texttt{mlbench} Benchmark}\label{sec:benchmark}

In this section, we present more details of the \texttt{mlbench} benchmark we constructed by harvesting and running winning code of Kaggle competitions.

\subsection{Kaggle Competitions}

Kaggle hosts various types of competitions for data scientists.
There are seven different competition categories, and we are particularly interested in the category ``Featured'' that aims to solve commercial, real-world machine learning problems.
For each competition, Kaggle provides a necessary description of its background, training and testing datasets, evaluation metric, and so on.
These competitions are online only for a while, and Kaggle allows participants to submit multiple times during that period.
Kaggle evaluates and provides a score for each submission (shown as a public leader board).
Participants can specify their final submissions before a competition ends, and a final, private leader board is available after the competition ends.
The winners are determined based on their rankings on the private leader board.
In this paper, we treat the top ten on the private leader board as ``winning code,'' and we look for the one ranked the highest among the top ten.

\subsection{Overview} \label{sec:mlbench:overview}

\begin{figure}[t]
\scriptsize
\centering
\begin{tabular}{ c | c  }
\hline
Statistics of Kaggle Competitions & Number \\ \hline
Total Competitions & 267 \\ 
Competitions with Winning Code & 41 \\ 
Competitions without Winning Code & 226  \\ \hline
\end{tabular}
\vskip -2ex
\caption{Statistics of Kaggle Competitions}
\label{tab:stats-1}
\vskip -3ex
\end{figure}

\texttt{mlbench} is curated from Kaggle competitions with or without winning code.
We describe the protocol of curating as follows.

\vspace{0.3em}
\noindent
{\underline{\bf Datasets from Winning Code.}}  
As shown in Figure~\ref{tab:stats-1}, we collected 267 Kaggle competitions in total and found winning code for 41 of these competitions.
We are unable to find winning code for the remaining 226 competitions.
Fortunately, the 41 competitions with available winning code already exhibit sufficient diversity to evaluate various aspects of machine learning clouds.
Figure~\ref{tab:stats-2} further summarizes the types of machine learning tasks covered by these 41 competitions with winning code.
Given the scope of study we stated in Section~\ref{sec:preliminaries:scope}, the 13 competitions that are binary classification tasks are the focus of our evaluation.

\begin{figure}
\scriptsize
\centering
\begin{tabular}{ c | c }
\hline
Tasks of Kaggle Competitions & Number \\ \hline
Binary Classification & 13 \\ 
Multi-class Classification & 14  \\ 
Regression & 9 \\ 
Others & 5 \\ \hline 
\end{tabular}
\vskip -2ex
\caption{Kaggle Competitions with Winning Code}
\label{tab:stats-2}
\vskip -5ex
\end{figure}

We then ran the winning code of the 13 competitions on Microsoft Azure for the purpose of extracting the features used by the winning code (recall Section~\ref{sec:preliminaries:methodology}).
We failed to run the winning code for ``Avito Context Ad Clicks''.
For ``Santander Customer Satisfaction'' and ``Higgs Boson Machine Learning Challenge'', the code cannot be finished on an Azure machine with a 16-core CPU and 112GB memory.
Therefore, there were 10 competitions for which we finished running the winning code successfully.
We further removed datasets whose
outputs are either three-dimensional features that cannot be supported by the machine learning clouds we studied
or features that cannot be extracted and saved successfully. 
Moreover, the winning code of ``KDD Cup 2014'' generated two sets of features --- it uses the ensemble method
with two models.
This results in 7 datasets with features extracted
by the winning code.

\vspace{0.3em}
\noindent
{\underline{\bf Beyond Winning Code.}}  
We also 
constructed datasets using the raw features from Kaggle (details in Section~\ref{sec:benchmark:dataset-details}), which results in 11 additional datasets.
Specifically, we include all binary competitions
ended by  July 2017 that (1) use AUC as
evaluation metric, (2) can be joined
by new users, (3) have datasets available for
download, (4) still allow for submission
and scoring, (6) do not contain
images, videos, and HTML files, and (5) whose
total size does not exceed Azure's limitation.\footnote{\url{https://docs.microsoft.com/en-us/azure/machine-learning/studio/faq}}

In total, \texttt{mlbench} contains 18 datasets
with 7 datasets having {\em both} features
produced by winning code and the raw features
provided by the competition. 
We summarize the statistics of the datasets in Figure~\ref{tab:stats-datasets}.
We can see a reasonable diversity across the datasets in terms of the size of the training set, the size of the testing set, and the number of features.
Moreover, the ratio between the sizes of the training set and testing set varies as well.
For example, D-VP has a testing set 10 times larger than the training set, which is quite different from the vanilla setting, where the training set is much larger.

\begin{figure}
\scriptsize
\centering
\begin{tabular}{l|rrrrr}

\hline
  Dataset &  Training Set &  Test Set &  \# Features & Training Size & Test Size \\
  \hline

 D-SCH-r &            86 &    119,748 &         410 &        0.3MB &     488MB \\
 D-PIS-r &          5,500 &      5,952 &          22 &         1.2MB &    1.29MB \\
  D-EG-r &          7,395 &      3,171 &          124 &          21MB &       9MB \\
 D-VPr-r &         10,506 &    116,293 &          53 &           2MB &      19MB \\
 D-AEA-r &         32,769 &     58,921 &           9 &        1.94MB &    3.71MB \\
 D-SCS-r &         76,020 &     75,818 &         369 &        56.6MB &    56.3MB \\
 D-SMR-r &        145,231 &    145,232 &        1,933 &         921MB &     921MB \\
 D-GMC-r &        150,000 &    101,503 &          10 &        7.21MB &    4.75MB \\
 D-HQC-r &        260,753 &    173,836 &         297 &         198MB &     131MB \\
 D-KDD-r &        619,326 &     44,772 &         139 &         571MB &    40.7MB \\
 D-PBV-r &       2,197,291 &    498,687 &          52 &         181MB &      76MB \\
 \hline
   D-SCH &            86 &    119,748 &         410 &         0.3MB &     488MB \\
    D-EG &          7,395 &      3,171 &          29 &        2.59MB &    1.35MB \\
    D-VP &         10,506 &    116,293 &          17 &         0.8MB &     9.1MB \\
   D-AEA &         32,769 &     58,921 &         135 &          54MB &      97MB \\
   D-PHY &         38,012 &    855,819 &          74 &        28.9MB &     859MB \\
    D-KDD2 &        131,329 &     44,772 &         100 &         105MB &      36MB \\
  D-KDD1 &        391,088 &     44,772 &         190 &         282MB &      32MB \\

\hline
\end{tabular}
\caption{Statistics of datasets. The ``-r'' in the names of  datasets indicates raw feature (see Section~\ref{sec:benchmark:dataset-details} for details).}
\label{tab:stats-datasets}
\vskip -6ex
\end{figure}

\subsection{Dataset Details}\label{sec:benchmark:dataset-details}

We present more details about the datasets listed in Figure~\ref{tab:stats-datasets}.
For each dataset, we first introduce the background of the corresponding Kaggle competition.
We then describe the features used by the winning code we found, which characterize the datasets themselves, as well as the models and algorithms it adopts.

\begin{itemize}

    \item {\bf MLSP2014-Schizophrenia Classification Challenge\\ (D-SCH and D-SCH-r)}: In this competition, multimodal features derived from brain magnetic resonance imaging (MRI) scans and labels of the training data are provided. The goal of this competition is to build machine learning models that can predict whether a person is a ``healthy control'' or ``schizophrenic patient'' in the testing data.\footnote{\scriptsize https://www.kaggle.com/c/mlsp-2014-mri}
    We use the winning code from Karolis Koncevicius~\cite{winningcode:schizophrenia}. Interestingly, the winning code uses the same features as the raw data provided by Kaggle. The algorithm it uses is distance weighted discrimination~\cite{DWD}. We abbreviate the constructed dataset as \textbf{D-SCH}. Another dataset is constructed using the raw data provided by Kaggle and is referred to as \textbf{D-SCH-r}, although \textbf{D-SCH} and \textbf{D-SCH-r} contain the same features in this particular case. 
    
    \item {\bf Influencers in Social Networks (D-PIS-r)}: In this competition, each data point describes the features extracted based on the Twitter activities of two individuals. In the training dataset, the data points are labelled to indicate which one of the two individuals is more influential on a social network. The goal of this competition is to predict the more influential individual from the given features of 2 individuals. \footnote{\scriptsize https://www.kaggle.com/c/predict-who-is-more-influential-in-a-social-network.} No winning code is available for this competition. We take the raw data provided by Kaggle and construct the dataset \textbf{D-PIS-r}.
    
    \item {\bf StumbleUpon Evergreen Classification Challenge \\ (D-EG and D-EG-r)}: In this competition, URLs and their corresponding raw contents are given. The goal of this competition is to build a classifier that can label a URL as either ``evergreen'' or ``ephemeral.''~\footnote{\scriptsize https://www.kaggle.com/c/stumbleupon}
    We use the winning code from Marco Lui~\cite{winningcode:evergreen}. It extracts features from raw HTML documents and uses only text-based features. The most important step is a stacking-based approach that combines the generated features~\cite{Lui2012}. The algorithm the winning code uses is logistic regression. Features used by this winning code are stored by extracting the input to the logistic regression classifier. We abbreviate this constructed dataset as \textbf{D-EG}. Similarly, we construct another dataset \textbf{D-EG-r} using the raw features.
     
    \item {\bf West-Nile Virus Prediction (D-VP and D-VP-r)}: In this competition, the participants are given weather, location, testing, and spraying data to predict whether or not West Nile Virus is present.~\footnote{\scriptsize https://www.kaggle.com/c/predict-west-nile-virus}
    We use the winning code from~\cite{winningcode:virusprediction}. For feature engineering, three new features are added to the dataset. In addition, another feature, ``NumMosquitos'' (indicating the number of mosquitos caught in a trap), exists in the training data but does not exist in the testing data. The author estimates this value for the testing data twice, and we take the average. 
    The predictions are initialized according to a normal distribution. Each prediction is then multiplied by various coefficients in several steps. These coefficients are obtained from other information related to the target (e.g., geographical information). The predictions are then normalized. We abbreviate this dataset as \textbf{D-VP}. The corresponding dataset using raw features is denoted as \textbf{D-VP-r}.

    \item {\bf Amazon.com-Employee Access Challenge (D-AEA \\ and D-AEA-r)}: The historical data that employees are allowed or denied to access resources over time is given. The goal of this competition is to create an algorithm that can predict approval/denial for an unseen employee.\footnote{\scriptsize https://www.kaggle.com/c/amazon-employee-access-challenge}
    We use the winning code from Owen Zhang~\cite{winningcode:amazon}. It first converts the original categorical features to numerical features. It then builds six models from subsets of the features as well as features obtained via post-processing (e.g., aggregation). The final prediction is generated by an ensemble of predictions from individual models. The algorithms it uses are GBM (generalized boosted regression modeling)~\cite{gbm}, random forest~\cite{random-forest}, extremely randomized trees~\cite{ert}, and glmnet (lasso and elastic-net regularized generalized linear models)~\cite{glmnet}. Features used by this winning code are stored by merging all features used by the models. We abbreviate this constructed dataset as \textbf{D-AEA}. Correspondingly, the dataset containing only the raw data from Kaggle is denoted as \textbf{D-AEA-r}. 
    
    \item {\bf Santander Customer Satisfaction (D-SCS-r)}: In this competition\footnote{\scriptsize https://www.kaggle.com/c/santander-customer-satisfaction}, the objective is to identify if a customer is unsatisfied with their experience in dealing with the Santander bank. A list of numeric features as well as a label are provided to the participants. There is no winning code available for this competition. We use the raw data provided by Kaggle to construct the dataset \textbf{D-SCS-r}.
    
    \item {\bf Springleaf Marketing Response (D-SMR-r)}: In this \\competition\footnote{\scriptsize https://www.kaggle.com/c/springleaf-marketing-response}, a large set of anonymized features describing a customer are provided in each entry of the training dataset. The goal of this competition is to use the features of the customer to predict whether s/he will respond to a direct mail offer. 
    No winning code is available for this competition. We therefore construct the dataset \textbf{D-SMR-r} using the raw data from Kaggle.
    
    \item {\bf Give me some credit (D-GMC-r)}: In this competition\footnote{\scriptsize https://www.kaggle.com/c/GiveMeSomeCredit}, the participants are asked to help a bank to predict the probability that a client will experience financial distress in the next two years. No winning code is available for this dataset. We take the raw data from Kaggle and denote the dataset as \textbf{D-GMC-r}. 
    
    \item {\bf Homesite Quote Conversion (D-HQC-r)}: In this competition\footnote{\scriptsize https://www.kaggle.com/c/homesite-quote-conversion}, the participants are asked to predict whether or not a customer will purchase the quoted product from an insurance company. The training data includes anonymized features covering information about the product, the client, the property going to be assured, and the location. 
    We use the raw data to create the dataset \textbf{D-HQC-r}. 

    \item {\bf KDD Cup 2014-Predicting Excitement (D-KDD1,  D-KDD2 and D-KDD-r)}: In this competition, the participants are asked to help DonorsChoose.org identify projects that are exceptionally exciting to the business, given all the related data about projects, donations, and so on.~\footnote{\scriptsize https://www.kaggle.com/c/kdd-cup-2014-predicting-excitement-at-donors-choose}
    We use the winning code from~\cite{winningcode:kddcup}. It builds two diverse feature sets based on raw features and generated features. These two diverse feature sets are then used to train two different models: gradient boosting regressor and gradient boosting machine. The final result is based on the ensemble of the two. We abbreviate these two constructed datasets as \textbf{D-KDD1} and \textbf{D-KDD2}. 
    As before, we also create a dataset that only contains the raw data, denoted as \textbf{D-KDD-r}. 

    \item {\bf Predicting Red Hat Business Value (D-PBV-r)}: The goal of this competition\footnote{\scriptsize https://www.kaggle.com/c/predicting-red-hat-business-value} is to identify customers with potential business value. To achieve this goal, records of customer activities are provided to the participants. 
    In addition, each customer is associated with a set of features. There is no winning code for this competition. 
    We construct the dataset \textbf{D-PBV-r} by joining the tables containing raw data describing the activities and the features of the customers.
    
    \item {\bf Flavours of Physics: Finding $\tau\to\mu\mu\mu$ (D-PHY)}: In this competition, the participants are given a list of collision events and their properties to predict whether $\tau\to 3\mu$ decay happens in a collision or not.\footnote{\scriptsize https://www.kaggle.com/c/flavours-of-physics}
    We use the winning code from Alexander V. Gramolin~\cite{winningcode:physics}. It designs new features based on original features. One original feature is not used because it prevents passing of the agreement test.\footnote{\scriptsize https://www.kaggle.com/c/flavours-of-physics/details/agreement-test}
    In addition, the winning code does not use all the training data.
    Regarding the algorithm, it uses only XGBoost~\cite{XGBoost}. It trains two different XGBoost models on different sets of features. The final result is an ensemble of results obtained by the two models.
    The combination of two independent classifiers enables it to pass the correlation test.\footnote{\scriptsize https://www.kaggle.com/c/flavours-of-physics/details/correlation-test} Features used by this winning code are stored by extracting the input to the models. Then the features taken into different models are merged and duplicated features are dropped. We abbreviate this constructed dataset as \textbf{D-PHY}.

\end{itemize}

There are missing values in the datasets D-PHY and D-KDD2. We replace the missing values in these two datasets with the average values of corresponding features and ``N/A'', respectively, for our experiments on Azure and
Amazon.

\section{Experimental Settings}\label{sec:settings}

We evaluate the declarative machine learning services provided by two major cloud vendors: Microsoft Azure Machine Learning Studio and Amazon Machine Learning. We will use {\bf Azure} and {\bf Amazon} as shorthand.

We first introduce the current APIs of {\bf Azure} and {\bf Amazon} and then all machine learning models they provide.

\subsection{Existing Cloud API}

Both {\bf Azure} and {\bf Amazon} start by asking users to upload their data, which can be in the form
of CSV files.
Users then specify the machine learning tasks they want to run on the cloud. 
However, {\bf Azure} and {\bf Amazon} offer different APIs, as illustrated below.

\begin{itemize}
    \item {\bf Azure} provides an API using which users specify the {\em types} of {\em machine learning models}, such as (1) logistic regression, (2) support vector machine, (3) decision tree, etc.
    For each type of model, {\bf Azure} provides a set of default hyper-parameters for users to use in an out-of-the-box manner. 
    {\bf Azure} also supports different ways of automatic hyper-parameter tuning and provides a default range of values to be searched for.
    \item {\bf Amazon} provides an API by which users specify the {\em types} of {\em machine learning tasks}, namely (1) binary classification, (2) multiclass classification, and (3) regression.
    For each type, {\bf Amazon} automatically chooses the type of machine learning models.
    For now, {\bf Amazon} always runs a logistic regression for binary classification~\cite{amazoncloud}.
    {\bf Amazon} further provides a set of default hyper-parameters for logistic regression, but users can also change these default values.
\end{itemize}

\begin{figure}
\tiny
\centering
\begin{tabular}{l|l|l|l}
\hline
   Platform & Model & \# Combinations & Tuned Parameters\\
\hline
\multirow{8}{3em}{{\bf Azure}} 
& \multirow{2}{3em}{C-AP} & \multirow{2}{2em}{6} & Learning Rate=\{0.1, 0.5, \textbf{1.0}\}\\
& &  & Maximum \# iterations=\{1, \textbf{10}\}\\
\cline{2-4}
& C-BPM & 1 & \# iterations=\{\textbf{30}\} \\
\cline{2-4}
& \multirow{4}{3em}{C-BDT} & \multirow{4}{2em}{180} & Maximum \# leaves per tree=\{2, 8, \textbf{20}, 32, 128\}\\
& & &Minimum \# samples per leaf=\{1, \textbf{10}, 50\}\\
& & &learning rate =\{0.025, 0.05, 0.1, \textbf{0.2}, 0.4\}\\
& & &\# trees=\{ 20, \textbf{100}, 500\} \\
\cline{2-4}
& \multirow{4}{3em}{C-DF} & \multirow{4}{2em}{81} & \# trees=\{1, \textbf{8}, 32\}\\
& & &Maximum depth=\{1, 16, \textbf{32}, 64\}\\
& & &\# Random splits per note=\{1, \textbf{128}, 1024\}\\
& & &Minimum \# samples per leaf=\{\textbf{1}, 4, 16\} \\
\cline{2-4}
& \multirow{4}{3em}{C-DJ} & \multirow{4}{2em}{81} & \# Decision DAGs=\{1, \textbf{8}, 32\}\\
& & &Max depth of decision DAGs=\{1, \textbf{32}, 16, 64\}\\
& & &Max width of decision DAGs=\{1, \textbf{128}, 1024\}\\
& & &\# optimizations per layer=\{1024, \textbf{2048}, 4096, 16384\}\\

\cline{2-4}

& \multirow{3}{3em}{C-NN} & \multirow{3}{2em}{12} & Learning Rate=\{\textbf{0.01}, 0.02, 0.04\} \\
& & & \# iterations=\{20, 40, 80, \textbf{100}, 160\} \\

\cline{2-4}
& \multirow{2}{3em}{C-SVM} & \multirow{2}{2em}{15} & \# iterations=\{\textbf{1}, 10, 100\} \\
& & & $\lambda=\{1*10^{-5,-4,\textbf{-3},-2,-1}\}$\\
\cline{2-4}

& \multirow{4}{3em}{C-LR} & \multirow{4}{2em}{72} & Optimization tolerance=$\{1*10^{\textbf{-7},-4}\}$\\
& & &L1 regularization weight=\{0, 0.01, 0.1, \textbf{1.0}\}\\
& & &L2 regularization weight=\{0.01, 0.1, \textbf{1.0}\}\\
& & &Memory size for L-BFGS=\{5, \textbf{20}, 50\}\\

\hline
{\bf Amazon} & C-LR & N/A & Amazon automatically tunes the learning rate \\
\hline
\end{tabular}
\vskip -2ex
\caption{Hyper-parameters tuned for each model. The default parameters are in bold font.}
\label{table:parameter}
\vskip -6ex
\end{figure}

\begin{figure}
\tiny
\centering
\begin{tabular}{l|rrrrrrrr}

\hline
Data Set &   C-AP & C-BPM &  C-BDT &   C-DF &   C-DJ &   C-LR &   C-NN &  C-SVM\\ 

\hline

  D-SCH-r &      6 &     9 &     280 &     12 &    130 &     15 &     11 &     11\\ 
  D-PIS-r &      6 &     5 &     639 &    178 &    551 &     16 &     30 &     16\\ 
   D-EG-r &     28 &    15 &   10,165 &    260 &    530 &    145 &   3,548 &    271\\ 
   D-VP-r &     32 &     8 &    3,105 &    821 &   3,342 &    135 &    824 &    304 \\ 
  D-AEA-r &     33 &    22 &   25,240 &   2,028 &   3,968 &    230 &    690 &    229 \\ 
  D-SCS-r &    126 &   286 &    6,198 &   2,816 &   7,119 &    791 &   1,583 &   1,433 \\ 
  D-SMR-r &  18,315 &  3,116 &     NA &  15,853 &  19,332 &    NA &    NA &    NA \\ 
  D-GMC-r &     27 &    17 &    4,146 &   4,425 &  10,635 &    206 &    501 &    208 \\ 
  D-HQC-r &   2,697 &   454 &   47,502 &  22,023 &    NA &  12,404 &  29,617 &  27,341 \\ 
  D-KDD-r &   4,290 &   906 &     NA &  40,816 &  63,185 &  37,328 &    NA &  39,716 \\ 
  D-PBV-r &  11,450 &   877 &     NA &    NA &    NA &  84,290 &    NA &    NA \\ 
  \hline
    D-SCH &      6 &     9 &     280 &     12 &    130 &     15 &     11 &     11 \\ 
     D-EG &      6 &    13 &     837 &    227 &    690 &     17 &     39 &     22 \\ 
     D-VP &     13 &    14 &    1,546 &    422 &   2,383 &     55 &    216 &    100 \\ 
    D-AEA &     25 &   241 &    3,495 &    780 &   3,413 &    210 &    323 &    193 \\ 
    D-PHY &     21 &   114 &    2,901 &    644 &   3,167 &    132 &    262 &    144 \\ 
   D-KDD2 &    550 &  1,450 &  365,256 &   6,675 &   9,339 &   7,228 &    NA &   6,191 \\ 
   D-KDD1 &    310 &  3,814 &   40,364 &  24,796 &  88,044 &   4,455 &   4,465 &   3,074 \\ 

\hline
\end{tabular}
\vspace{-2em}
\caption{Total training time (seconds) on Azure with hyper-parameter tuning (HPT).}
\label{table:performance}
\vskip -4ex
\end{figure}

\begin{figure}
\tiny
\centering
\begin{tabular}{l|rrrrrrrr}

\hline
Dataset &  C-AP & C-BPM & C-BDT & C-DF & C-DJ & C-LR &   C-NN & C-SVM\\ 

\hline

  D-SCH-r &     4 &     6 &     5 &    3 &    3 &    3 &      4 &     4\\ 
  D-PIS-r &     3 &     5 &     6 &    7 &    4 &    4 &      6 &     4\\ 
   D-EG-r &     7 &    15 &     9 &    6 &    4 &    4 &    247 &     4\\ 
   D-VP-r &     6 &     8 &     7 &    7 &    7 &    5 &     69 &     5\\ 
  D-AEA-r &     5 &    22 &    12 &    6 &    9 &    5 &     58 &     5 \\ 
  D-SCS-r &    10 &   286 &    10 &   18 &   11 &    6 &    109 &     6 \\ 
  D-SMR-r &  1,048 &  3,116 &  1,148 &   45 &   50 &  693 &    NA &   328 \\ 
  D-GMC-r &     5 &    17 &    10 &   59 &   14 &    4 &     42 &     4 \\ 
  D-HQC-r &   189 &   454 &    58 &  113 &  143 &   57 &   2,254 &    54 \\ 
  D-KDD-r &   222 &   906 &   905 &   48 &   99 &  122 &    NA &    88 \\ 
  D-PBV-r &   513 &   877 &   439 &  160 &  788 &  229 &    NA &   112\\ 
  \hline
    D-SCH &     4 &     4 &     5 &    3 &    3 &    4 &      4 &     4 \\ 
     D-EG &     4 &     5 &     5 &    7 &    4 &    4 &      6 &     4 \\ 
     D-VP &     5 &     5 &     5 &    7 &    6 &    4 &     17 &     4 \\ 
    D-AEA &     6 &    54 &     9 &   10 &    8 &    5 &     24 &     5\\ 
    D-PHY &     5 &    23 &     8 &    9 &    8 &    5 &     22 &     5 \\ 
   D-KDD2 &    50 &   339 &   146 &   14 &   22 &   27 &    NA &    20 \\ 
   D-KDD1 &   183 &   412 &   161 &   41 &  117 &   73 &  46,830 &    55\\ 

\hline
\end{tabular}

\vspace{-2em}
\caption{Average HPT training time (seconds) on Azure.}
\label{table:performance-average}
\vskip -6ex
\end{figure}

\vspace{-1em}
\subsection{Machine Learning Models}

In the following, we give a brief description of the machine learning models provided by {\bf Azure} and {\bf Amazon}.

\begin{itemize}
    \item {\bf Two-Class Averaged Perceptron (C-AP)}: It is a linear classifier and can be thought of as a simplified neural network: there is only one layer between input and output.\footnote{\scriptsize https://msdn.microsoft.com/en-us/library/azure/dn906036.aspx}
    
    \item {\bf Two-Class Bayes Point Machine (C-BPM)}: This is a Bayesian classification model, which is not prone to overfitting. The Bayes point is the average classifier that efficiently approximates the theoretically optimal Bayesian average of several linear classifiers (in terms of generalization performance)~\cite{BPM}.
    
    \item {\bf Two-Class Boosted Decision Tree (C-BDT)}: Boosting is a well-known ensemble algorithm that combines weak learners to form a stronger learner (e.g., AdaBoost~\cite{FreundS97}). The boosted decision tree is an ensemble method that constructs a series of decision trees~\cite{Quinlan86}. Except for the first tree, each of the remaining trees is constructed by correcting the prediction error of the previous one.
    The final model is an ensemble of all constructed trees.
    
    \item {\bf Two-Class Decision Forests (C-DF)}: This classifier is based on random decision forests~\cite{Ho95}. Specifically, it constructs multiple decision trees that vote on the most popular output class.\footnote{\scriptsize https://msdn.microsoft.com/en-us/library/azure/dn906008.aspx}
    
    \item {\bf Two-class Decision Jungle (C-DJ)}: This is an ensemble of rooted decision directed acyclic graphs (DAGs). In conventional decision trees, only one path is allowed from the root to a leaf. In contrast, a DAG in a decision jungle allows multiple paths from the root to a leaf~\cite{ShottonSKNWC13}.
    
    \item {\bf Two-Class Logistic Regression (C-LR)}: This is a classic classifier that predicts the probability
    of an instance by fitting a logistic function.\footnote{\scriptsize https://msdn.microsoft.com/en-us/library/azure/dn905994.aspx} It is also the only classifier that {\bf Amazon} supports.

    \item {\bf Two-Class Neural Network (C-NN)}: Neural networks are bio-inspired algorithms that are loosely analogous to the observed behavior of a biological brain's axons~\cite{NN}. Specifically, the input layer (representing input data) and the output layer (representing answers) are connected by layers of weighted edges and nodes, which encode the so-called activation functions.\footnote{\scriptsize https://msdn.microsoft.com/en-us/library/azure/dn905947.aspx}
    
    \item {\bf Two-Class Support Vector Machine (C-SVM)}: SVM is another well-known classifier~\cite{SVM}. It works by separating the data with the ``maximum-margin'' hyperplane.\footnote{\scriptsize https://msdn.microsoft.com/en-us/library/azure/dn905835.aspx}
\end{itemize}

\subsection{Hyper-parameter Tuning}

\begin{figure*}[t]
\centering

\includegraphics[width=1.6\columnwidth]{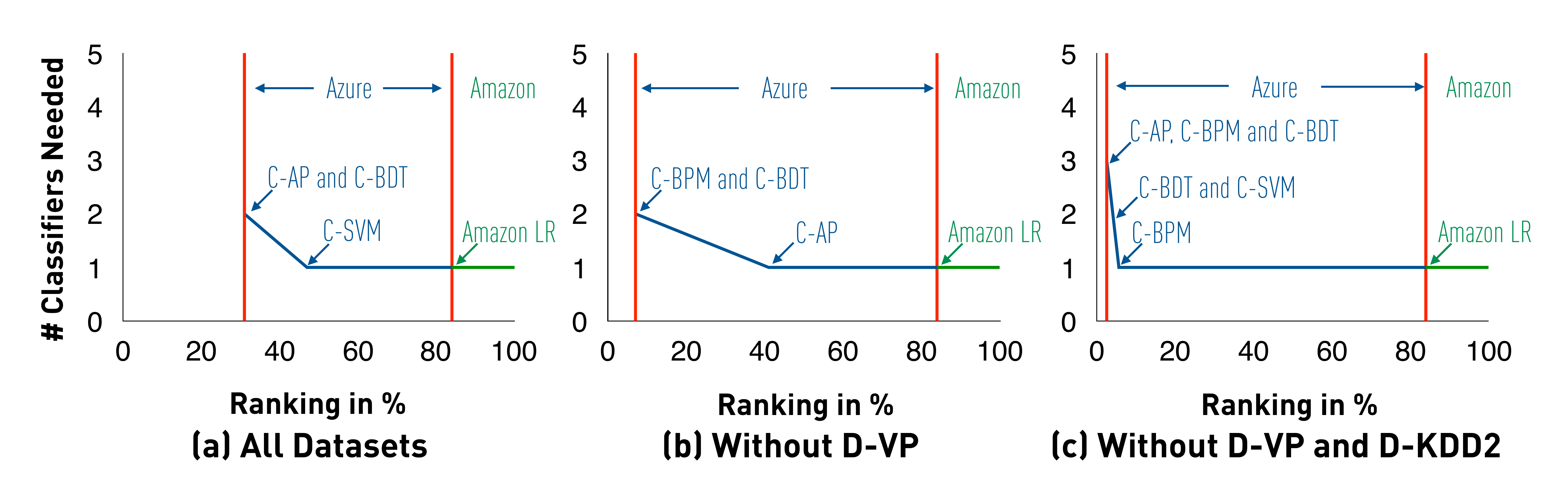}
\vskip -4ex
\caption{Capacity and universality of machine learning clouds as quality tolerance various.}
\label{fig:main}
\vskip -2ex
\end{figure*}

\begin{figure*}
\scriptsize
\centering
\begin{tabular}{l|rr|rrrrrrrr|r}

\hline
   Dataset & Leaderboard\#1 & WinningCode &         C-AP &        C-BPM &        C-BDT &         C-DF &         C-DJ &         C-LR &         C-NN &        C-SVM &          Amazon \\
\hline
    D-SCH-r &      0.93 (1) &    0.91 (3)    & 0.88 (37) &     0.92 (3) &   0.69 (279) &   0.68 (281) &   0.82 (165) &    0.89 (22) &    0.88 (30) &    0.90 (8) &  0.74 (264) \\
 D-PIS-r &      0.88 (1) &             &   0.76 (127) &   0.78 (118) &    0.87 (27) &    0.86 (62) &    0.87 (48) &   0.78 (118) &   0.80 (114) &   0.77 (120) &    0.86 (66) \\
  D-EG-r &      0.89 (1) &   0.89 (4)    &   0.86 (388) &   0.86 (381) &   0.86 (374) &   0.86 (382) &   0.86 (388) &   0.86 (388) &   0.86 (388) &   0.85 (400) &    NA (NA) \\
  D-VP-r &      0.86 (1) &   0.86 (1)   &  0.63 (1105) &  0.64 (1083) &   0.69 (840) &  0.63 (1100) &  0.60 (1137) &   0.67 (928) &  0.66 (1023) &  0.57 (1157) &   0.66 (989) \\
 D-AEA-r &      0.92 (1) &   0.92 (2)   &   0.87 (834) &   0.84 (932) &   0.87 (802) &   0.86 (889) &  0.79 (1049) &   0.88 (737) &   0.87 (798) &   0.86 (869) &   0.85 (901) \\
 D-SCS-r &      0.83 (1) &  0.83 (3)   &  0.75 (4128) &  0.77 (4038) &  0.82 (3084) &  0.82 (3432) &  0.82 (3240) &  0.78 (4019) &  0.79 (3916) &  0.71 (4302) &  0.81 (3691) \\
 D-SMR-r &      0.80 (1) &             &  0.74 (1826) &  0.70 (1898) &    NA (NA) &  0.73 (1842) &  0.71 (1888) &    NA (NA) &    NA (NA) &    NA (NA) &    NA (NA) \\
 D-GMC-r &      0.87 (1) &             &   0.74 (810) &   0.70 (820) &   0.87 (137) &   0.86 (505) &   0.87 (139) &   0.70 (837) &   0.83 (713) &   0.81 (750) &   0.86 (544) \\
 D-HQC-r &      0.97 (1) &             &  0.94 (1431) &  0.93 (1476) &   0.97 (977) &  0.96 (1206) &    NA (NA) &  0.94 (1429) &  0.95 (1335) &  0.94 (1463) &  0.96 (1294) \\
 D-KDD-r &      0.68 (1) &    0.67 (2)   &   0.59 (130) &   0.55 (314) &    NA (NA) &   0.56 (297) &   0.54 (373) &    0.60 (77) &    NA (NA) &   0.54 (363) &   0.58 (152) \\
 D-PBV-r &      1.00 (1) &             &  0.95 (1651) &  0.98 (1498) &    NA (NA) &    NA (NA) &    NA (NA) &  0.97 (1529) &    NA (NA) &    NA (NA) &  0.96 (1612) \\
 
 \hline
 
  
    D-SCH &      0.93 (1) &    0.91 (3)    & 0.88 (37) &     0.92 (3) &   0.69 (279) &   0.68 (281) &   0.82 (165) &    0.89 (22) &    0.88 (30) &    0.90 (8) &  0.74 (264) \\
   D-EG &      0.89 (1) &   0.89 (4)  &     0.89 (2) &     0.89 (3) &    0.88 (54) &   0.88 (278) &   0.88 (291) &     0.89 (2) &     0.89 (4) &     0.89 (2) &   0.88 (326) \\
    D-VP &      0.86 (1) &  0.86 (1)   &   0.65 (1046) &  0.70 (734) &   0.76 (408) &   0.74 (473) &   0.72 (568) &   0.71 (648) &   0.74 (496) &   0.72 (613) &   0.68 (923) \\
   D-AEA &      0.92 (1) &  0.92 (2)   &    0.91 (75) &    0.91 (94) &    0.92 (29) &   0.90 (147) &    0.91 (63) &    0.91 (72) &    0.91 (72) &   0.90 (401) &   0.90 (361) \\
   D-PHY &      1.00 (1) &  1.00 (2)   &    NA (NA) &    NA (NA) &    NA (NA) &    NA (NA) &    NA (NA) &    NA (NA) &    NA (NA) &    NA (NA) &    NA (NA) \\
     D-KDD2 &      0.68 (1) &  0.67 (2)   &   0.58 (192) &   0.51 (399) &    0.62 (33) &    0.61 (53) &    0.60 (72) &   0.58 (198) &    NA (NA) &   0.57 (206) &    0.60 (65) \\
  D-KDD1 &      0.68 (1) &  0.67 (2)   &    0.65 (12) &    0.64 (25) &    0.64 (26) &    0.62 (40) &    0.64 (25) &   0.65 (12) &    0.60 (68) &   0.64 (20) &    0.63 (29) \\
\hline

\end{tabular}

\vskip -2ex
\caption{Area under curve (AUC) and rankings on the private leader board of Kaggle for various datasets and models. The
results for public leader board are similar. The winning code of KDD is an ensemble of the classifiers trained from both D-KDD1 and D-KDD2.}
\label{tab:quality}
\vskip -4ex
\end{figure*}

Each machine learning algorithm consists of a set of hyper-parameters
to tune. The methodology we use in this paper is to rely on the
{\em default} tuning procedure provided by the machine learning cloud.
Figure~\ref{table:parameter} summarizes the hyper-parameters provided by 
the machine learning clouds. For each machine learning
model, we conduct an exhaustive grid search on all possible parameter combinations.

Because {\bf Amazon} only has the option of logistic regression and automatically tunes the learning rate, we only tuned hyper-parameters for models provided by {\bf Azure}.
We performed hyper-parameter tuning in an exhaustive manner: for each combination of hyper-parameter values in the whole search space, we ran the model based on that setting.
The best hyper-parameter is then selected based on the AUC score obtained
with five-fold cross validation. AUC is the evaluation metric used
by all Kaggle competitions we included in \texttt{mlbench}.
The third column in Figure~\ref{table:parameter} presents the number of hyper-parameter combinations in the search space for each model.
For example, C-AP employs two hyper-parameter knobs, ``learning rate'' and ``maximum number of iterations,'' with three and two alternative values.
As a result, there are six combinations in total.

Just for completeness, we report the time spent on tuning hyper-parameters for {\bf Azure} models.
Figure~\ref{table:performance} presents the total time of trying all hyper-parameter combinations for each model and dataset. Figure~\ref{table:performance-average} further reports the average time for one hyper-parameter combination.

\section{Results on Winning Features}\label{sec:results}

We first evaluated the performance of {\bf Azure} and {\bf Amazon} {\em assuming users have already conducted
feature engineering and only use the machine learning
cloud as a declarative execution engine of 
machine learning models}. Our analysis in this section will mainly focus on the seven datasets where winning code is available (i.e., the datasets in Figure~\ref{tab:stats-datasets} without the `-r' suffix).
We will discuss the cases when raw features are
used in Section~\ref{sec:results:feature-engineering}.
For each dataset and model, we run {\bf Azure} and {\bf Amazon} for at most 24 hours.

\subsection{Capability and Universality}

We first report the performance of {\bf Azure} and {\bf Amazon}, based on the \emph{capacity} and \emph{universality} metrics defined in Section~\ref{sec:preliminaries:metric}.
Figure~\ref{fig:main} presents the result.

In Figure~\ref{fig:main}(a), the $x$-axis represents the quality tolerance, whereas the $y$-axis represents the number of models required if a machine learning cloud can achieve a certain tolerance level $\tau$ for all seven datasets (i.e., a $\tau$-university of seven). The minimum $\tau$ shown in Figure~\ref{fig:main}(a) then implies the capacity of a machine learning cloud. We observe that the capacity of {\bf Azure} is around 31 (i.e., $c(\text{Azure})=31$), whereas the capacity of {\bf Amazon} is around 83 (i.e., $c(\text{Amazon})=83$).
Under this measurement, state-of-the-art machine learning clouds are far from competitive than deliberate machine learning models designed manually: With the goal of meeting a $\tau$-university of seven, $\tau$ can only be as small as 31 for {\bf Azure} (and 83 for {\bf Amazon}). In other words, in at least one Kaggle competition (among the seven), {\bf Azure} is ranked outside the top 30\%, whereas {\bf Amazon} is ranked outside the top 80\% on the leader board.

However, we note that this might be a distorted picture given the existence of ``outliers.''
In Figure~\ref{fig:main}(b) and~\ref{fig:main}(c), we further present results by excluding the datasets D-VP and D-KDD2. 
Although the capacity of {\bf Amazon} remains the same, the capacity of {\bf Azure} improves dramatically:
$c(\text{Azure})$ drops to 7 by excluding D-VP and further drops to 5 by excluding D-KDD2, which suggests that {\bf Azure} can be ranked within the top 10\% or even the top 5\% in most of the Kaggle competitions considered.

\subsubsection{Breakdown and Analysis}

We next take a closer look at how the machine learning clouds perform in individual competitions.
Figure~\ref{tab:quality} reports the details of the AUC 
of different models on different datasets.
The number in parentheses next to the AUC is the rank (of this AUC) on the leader board.
We note that not every winning code we found is top-ranked.
Often, the top-ranked code is not available, and in this case, we seek the next available winning code (among the top 10) on the leader board. We have several interesting observations.

\paragraph*{Diversity of models is beneficial.} An obvious difference between {\bf Azure} and {\bf Amazon} is that {\bf Azure} provides more alternative models than {\bf Amazon}. While the reason for {\bf Amazon} to provide only logistic regression as the available model is unclear, the results presented in Figure~\ref{tab:quality} suggest that the additional models provided by {\bf Azure} do help. In more detail, Figure~\ref{tab:capacity} compares the capacity of {\bf Azure} and {\bf Amazon} on different datasets. We observe that {\bf Azure} always wins over {\bf Amazon} in terms of capacity, often by a large margin. The capacity of {\bf Azure} over all the datasets is 31.24 (6.99 if excluding D-VP and 2.54 if further excluding D-KDD2) versus 84.34 of {\bf Amazon}, as shown in Figure~\ref{fig:main}.

\begin{figure}[t]
\scriptsize
\centering
\begin{tabular}{llrr}
\hline
Dataset & {\bf Azure} (Model) & {\bf Amazon} & Winning\\
\hline
D-SCH (313) & 0.96 (C-BPM) & 84.34 & 0.96\\
D-EG (625) & 0.32 (C-AP) & 52.16 & 0.64\\
D-VP (1306) & 31.24 (C-BDT) & 70.67 & 0.08\\
D-AEA (1687) & 1.72 (C-BDT) & 21.40 & 0.12\\
D-KDD2 (472) & 6.99 (C-BDT) & 13.77 & 0.42\\
D-KDD1 (472) & 2.54 (C-LR) & 6.14 & 0.42\\

\hline
\end{tabular}
\vskip -2ex
\caption{Capacity of {\bf Azure} and {\bf Amazon} on different datasets (i.e., Kaggle competitions).}
\label{tab:capacity}
\vskip -6ex
\end{figure}

\begin{figure}
\scriptsize
\centering
\begin{tabular}{lrrr}
\hline
Dataset & {\bf Azure} (C-LR) & {\bf Amazon} & Winning\\
\hline
D-SCH (313) & 7.03 & 84.34 & 0.96\\
D-EG (625) & 0.32 & 52.16 & 0.64\\
D-VP (1306) & 49.62 & 70.67 & 0.08\\
D-AEA (1687) & 4.27 & 21.40 & 0.12\\
D-KDD2 (472) & 41.95 & 13.77 & 0.42\\
D-KDD1 (472) & 2.54 & 6.14 & 0.42\\

\hline
\end{tabular}
\vskip -2ex
\caption{Capacity of the logistic regression model (C-LR) from {\bf Azure} and {\bf Amazon} on different datasets.}
\label{tab:capacity-lr}
\vskip -6ex
\end{figure}

\paragraph*{Model selection is necessary.} For a given dataset, the variation in terms of prediction quality is quite large across different models. For example, by using the models provided by {\bf Azure} on the dataset ``D-SCH,'' the rank varies from 3 (as good as the winning code we found) to 281 (ranked at the bottom 10\% of 313 Kaggle competition participants). This makes model selection a difficult job for {\bf Azure} users. ({\bf Amazon} users do not have this problem, as logistic regression is their only option.)

\paragraph*{Hyperparameter tuning makes a difference
for a single model.} Both {\bf Azure} and {\bf Amazon} provide logistic regression. The difference is that {\bf Azure} provides more knobs for hyper-parameter tuning (recall Figure~\ref{table:parameter}). Figure~\ref{tab:capacity-lr} compares the capacity of the logistic regression model (``C-LR'') provided by {\bf Azure} and {\bf Amazon}. {\bf Azure} wins on most of the datasets, perhaps due to more systematic hyper-parameter tuning. (We do not know how {\bf Amazon} tunes the learning rate for logistic regression.) However, there is no free lunch: Hyper-parameter tuning is time-consuming (recall Figures~\ref{table:performance} and~\ref{table:performance-average}). We will analyze 
the impact on hyperparameter tuning across all models and datasets in Section~\ref{sec:results:feature-engineering} with Figure~\ref{tab:hpt}.

\subsection{Model Selection}

The previous section gives an overview of the performance of {\bf Azure} and {\bf Amazon} in terms of their capacity and universality.
However, although we observe that the additional models provided by {\bf Azure} significantly improve performance, model selection and hyper-parameter tuning become new challenges.

From the user's perspective, there is then a natural question: {\em Given a machine-learning task, which model should a user choose (for good performance)?}
The answer depends on (1) the capacity of the models, (2) the time the user is willing to spend on parameter tuning and training, and (3) the user's quality tolerance level.

In the following, we study the trade-off between these factors.
Our goal is not to give a definitive conclusion, which is in general impossible given the variety of machine-learning tasks and models. 
Rather, by presenting the results observed in our study, we hope we can give some insights into what is going on in reality to help users come up with their own recipes.

\subsubsection{Linear vs. Nonlinear Models}

\begin{figure}
\centering
    \includegraphics[trim=4 8 4 4, clip,width=1\columnwidth]{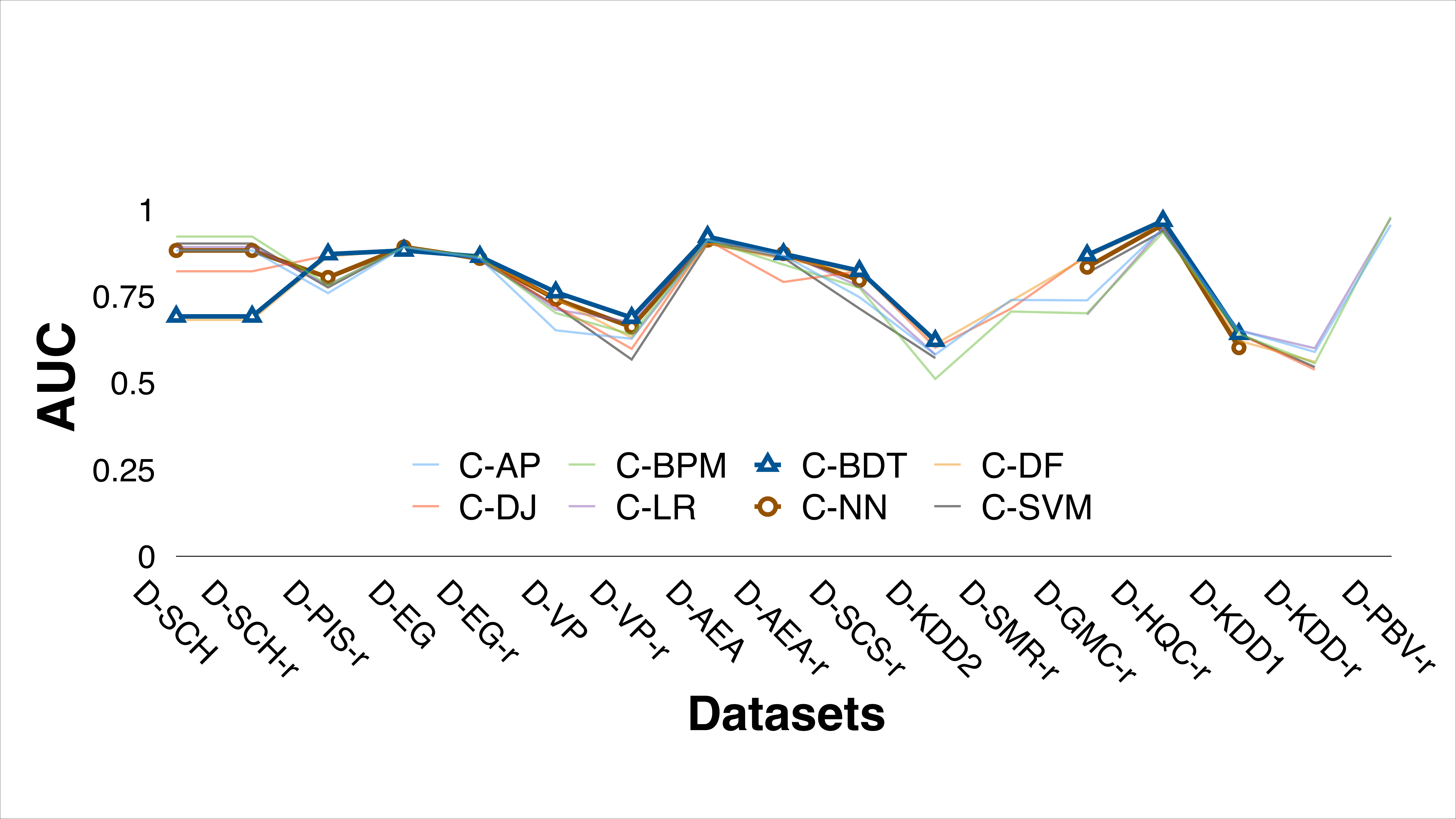}   
    \vskip -4ex
\caption{Quality of different models.}
\label{fig:quality}
\vskip -2ex
\end{figure}

In Figure~\ref{fig:main}, we have incrementally noted the models we need to include to improve the capacity of {\bf Azure} (with respect to a given universality).
Clearly, we find that nonlinear classifiers (e.g., C-BDT, C-NN, etc.) are the driving force that propels the improvement.
It is then an interesting question to investigate where the improvement indeed comes from.
We further compare the AUC of the models over different datasets in Figure~\ref{fig:quality}, based on the raw data in Figure~\ref{tab:quality}.

We observe that nonlinear models (e.g., C-BDT) dominate linear models (e.g., C-SVM) as the dataset size increases.
This is intuitive: Nonlinear models are more complicated than linear models in terms of the size of hypothesis space.
However, nonlinear models are more likely to suffer from overfitting on small datasets (e.g., the smallest dataset D-SCH in Figure~\ref{fig:quality}).

\begin{figure}
\centering
    \includegraphics[trim=8 8 4 4, clip,width=1\columnwidth]{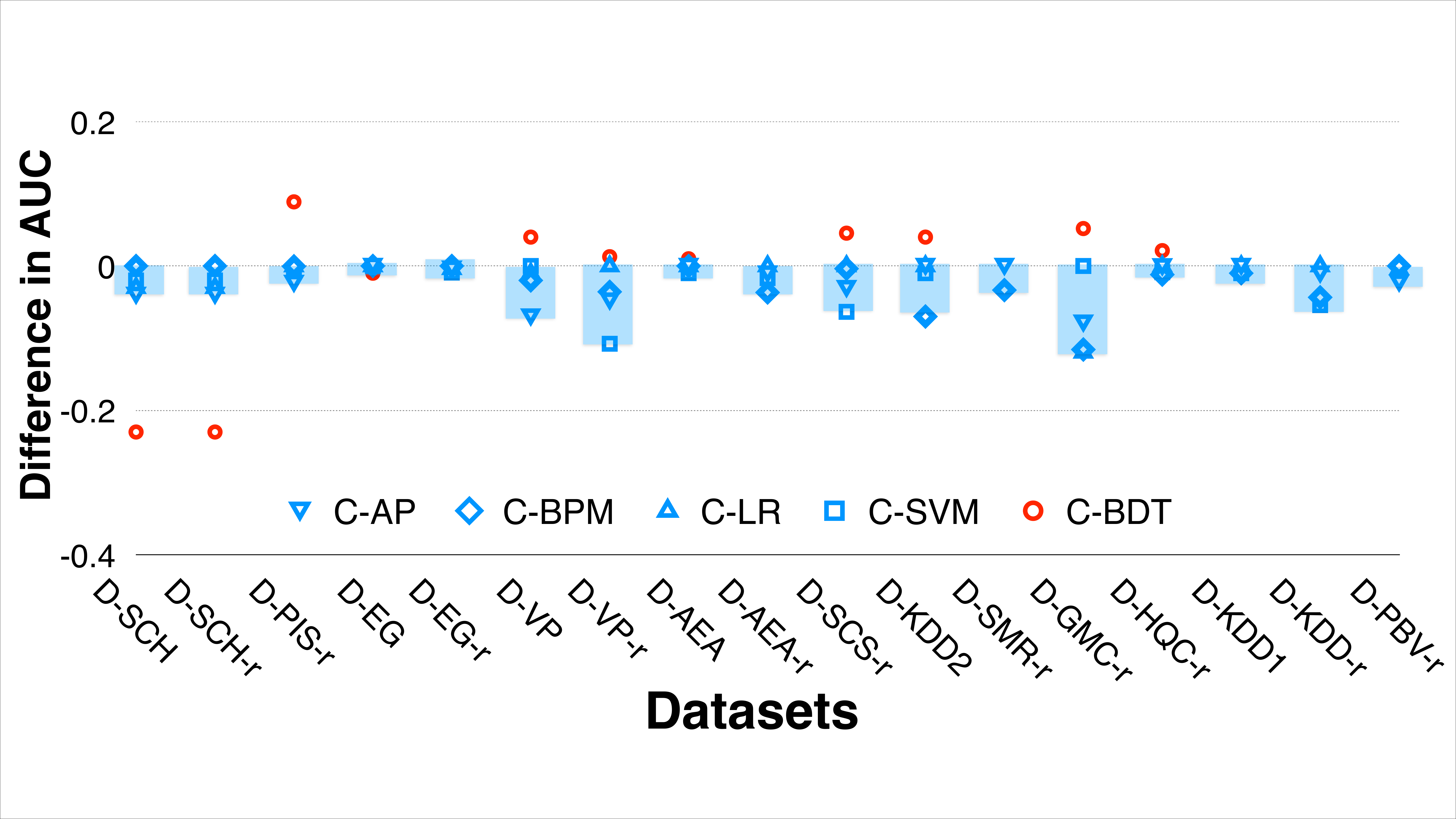}   
\vskip -4ex
\caption{BDT vs. Linear Classifiers.}
\label{fig:bdtvslin}
\vskip -2ex
\end{figure}

Figure~\ref{fig:bdtvslin} further presents a zoomed-in comparison between C-BDT, the dominant nonlinear classifier, and the linear models C-AP, C-BPM, C-SVM, and C-LR.
The $y$-axis represents the difference in terms of AUC between a model and the best linear model.
For example, the best linear model on the dataset D-SCH is C-BPM with an AUC of 0.92, whereas the best linear model on the dataset D-EG is C-SVM with an AUC of 0.89 (see Figure~\ref{tab:quality}).
Linear models often perform similarly regardless of dataset size: There is apparently a hit-or-miss pattern for linear models; namely, either the actual hypothesis falls into the linear space or it does not.
As a result, there is often no big difference in terms of prediction quality between linear models: If users believe that linear models are sufficient for a learning task, they can focus on reducing the training time rather than picking which model to use.

Based on these observations, our first empirical rule for model selection on machine-learning clouds is as follows:
\begin{observation}\label{observation:rule:size}
To maximize quality on Kaggle, use a nonlinear model
whenever it can scale and the dataset is not too small.
\end{observation}

\subsubsection{Training Time vs. Prediction Quality}

As we have mentioned, there is an apparent trade-off between the prediction quality of a model and the training time required for that model.
More sophisticated models usually have more knobs to tune and therefore need more time for training.
Given that nonlinear models in general outperform linear models on large datasets, it is worth further investigating the trade-off between their training time and prediction quality.

\begin{figure*}[t]
\centering
\includegraphics[width=1.8\columnwidth]{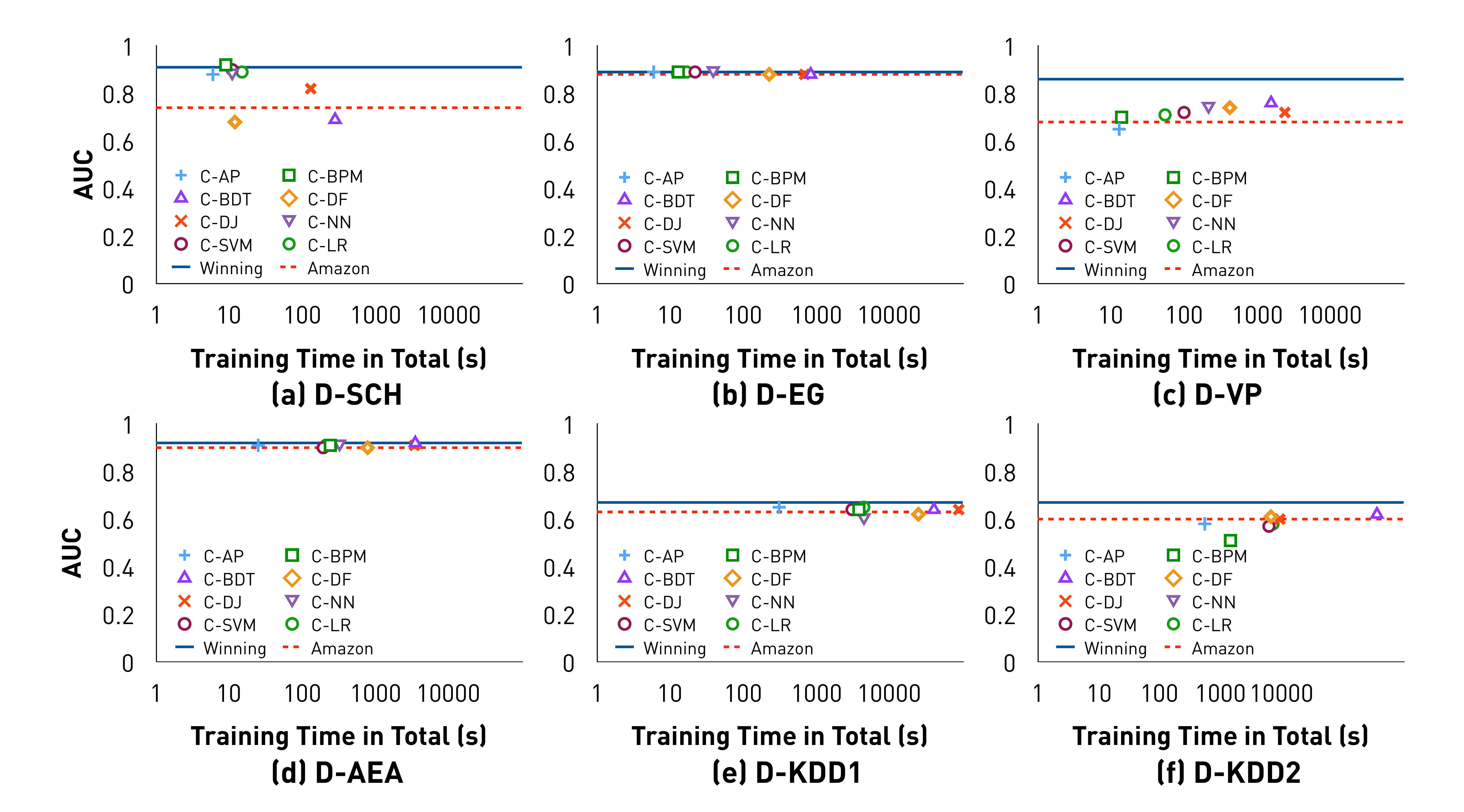}
\vskip -2ex
\caption{Tradeoff between prediction quality (AUC) and {\em total} training time. The blue line represents the AUC of the winning code, and the red line represents the AUC of logistic regression (C-LR) on {\bf Amazon}.}
\label{fig:trade-absolute}
\vskip -2ex
\end{figure*}

\begin{figure*}
\centering
\includegraphics[width=1.8\columnwidth]{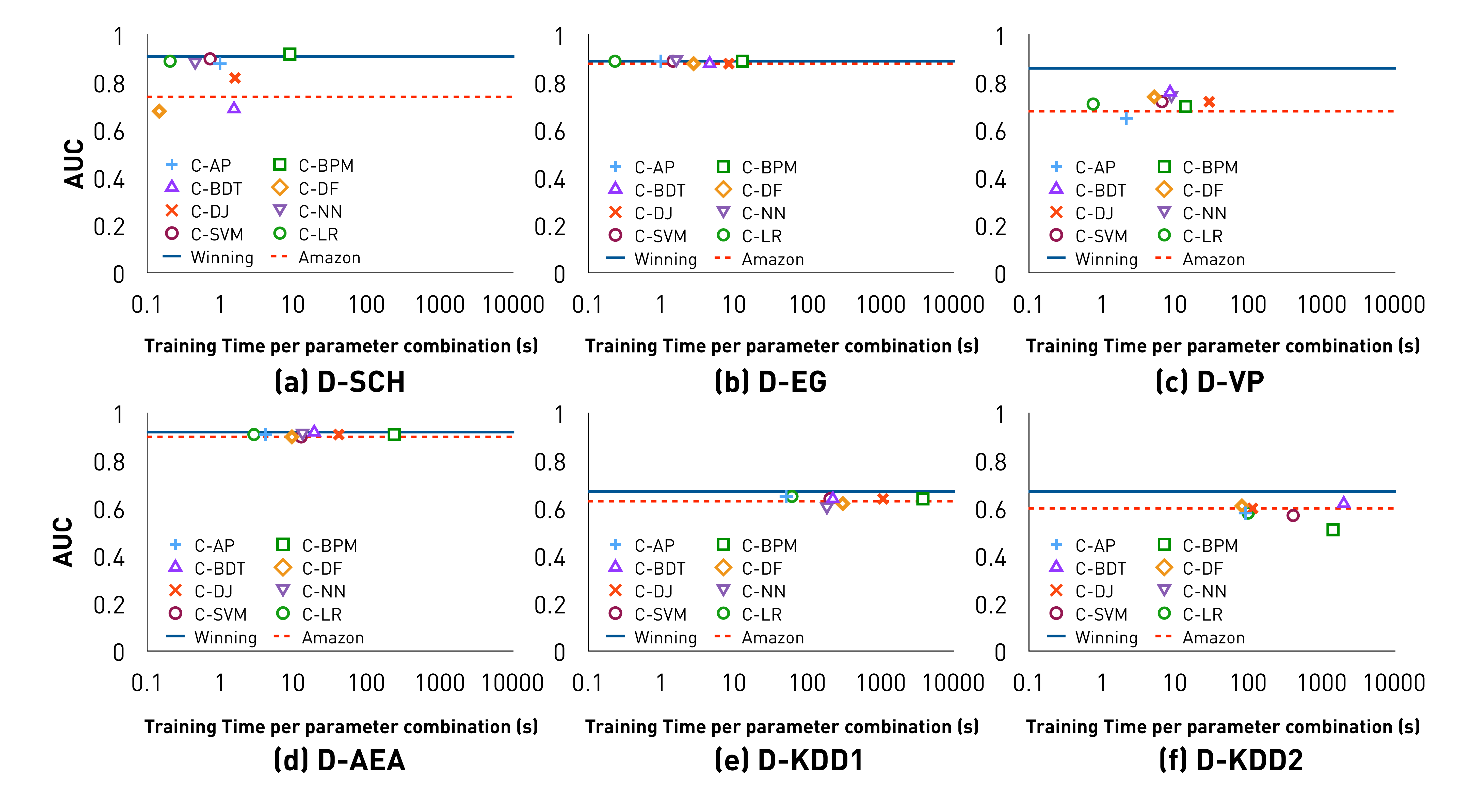}
\vskip -2ex
\caption{Tradeoff between prediction quality (AUC) and {\em average} training time per parameter. The blue line represents the AUC of the winning code and the red line represents the AUC of logistic regression (C-LR) on {\bf Amazon}.}
\label{fig:trade-average}
\vskip -2ex
\end{figure*}

We summarize the comparison result in Figures~\ref{fig:trade-absolute} and~\ref{fig:trade-average}.
Figure~\ref{fig:trade-absolute} presents the trade-off between the prediction quality and the total training time on hyper-parameter tuning, whereas Figure~\ref{fig:trade-average} presents the trade-off in the average sense (i.e., with respect to the average time spent on training a model under a specific hyper-parameter setting).
For ease of exposition, we order the models by their training time along the $x$-axis.
We also include linear models in our comparison for completeness.

In each plot of Figures~\ref{fig:trade-absolute} and~\ref{fig:trade-average}, the blue horizontal line represents the AUC of the winning code and the red horizontal line represents the AUC of (the logistic regression model provided by) {\bf Amazon}, whereas the scattered points present the AUC of {\bf Azure} models.
We have noted that the choice of linear versus nonlinear models can make a difference.
However, one interesting phenomenon we observe is that the choice within each category seems not so important; i.e., the prediction quality of different nonlinear models is similar.
Although this is understandable for linear models, it is a bit surprising for nonlinear models.
One reason for this is that most of the nonlinear models provided by {\bf Azure} are based on decision trees (C-BDT, C-DJ, and C-DF).
Moreover, more training time does not always lead to better prediction quality.
For example, in Figure~\ref{fig:trade-average}, the average training time of C-DJ is significantly longer than the others over the dataset D-KDD1. (Note that the $x$-axis is at the logarithmic scale.)
However, it is outperformed by even linear models such as C-AP, and its prediction quality is very close to that of C-BDT.

\begin{figure}
\scriptsize
\centering
\begin{tabular}{c|r|r}
\hline
Dataset & Winning AUC & 10\% AUC \\ \hline
D-SCH & 0.91 & 0.88 \\ 
D-EG & 0.89 & 0.88 \\ 
D-VP & 0.86 & 0.79 \\ 
D-AEA & 0.92 & 0.90 \\ 
D-KDD2 & \multirow{2}{*}{0.67} & \multirow{2}{*}{0.62}\\ 
D-KDD1 & &  \\ 
\hline
\end{tabular}
\vskip -2ex
\caption{AUC corresponding to the high tolerance regime of different Kaggle competitions.}
\label{tab:high-tolerance}
\vskip -6ex
\end{figure}

Based on these observations, our second empirical rule for model selection on machine-learning clouds is:
\begin{observation}\label{observation:rule:model}
To maximize efficiency,
within each category (linear or nonlinear), pick the one with the shortest training time.
\end{observation}
Considering the average training time presented in Figure~\ref{fig:trade-average}, C-AP is the choice among linear models, whereas C-BDT is the choice among nonlinear models.

\subsubsection{Quality Tolerance Regime}

We emphasize that the rules we presented in Observations~\ref{observation:rule:size} and~\ref{observation:rule:model} are purely empirical.
So far, we have looked at the model selection problem from only two of the three respects, i.e., the capacity of the models and the training time they require.
We now investigate the third respect: the user's quality tolerance.
This is a more fundamental and subtle point: A certain quality tolerance may not even be achievable for a machine learning task given the current capacity of machine learning clouds. (For example, we have seen that neither {\bf Azure} nor {\bf Amazon} can achieve even a quality tolerance of 30 on D-VP.)

To avoid oversimplifying the problem, we define the concept of {\em quality tolerance regime} based on the capacity of the machine learning clouds we currently observe:
\begin{itemize}
    \item {\em Low tolerance regime}. Corresponds to the case when the quality tolerance is below 1.\footnote{\scriptsize That is, when users can only be satisfied by winning the Kaggle competition or being ranked among the top 1\%.}
    \item {\em Middle tolerance regime}. Corresponds to the case when the quality tolerance is between 1 and 5.
    \item {\em High tolerance regime}. Corresponds to the case when the quality tolerance is between 5 and 10.
\end{itemize}

\vspace{-0.5em}
To give some sense of how well a model must perform to meet the tolerance regimes, in Figure~\ref{tab:high-tolerance}, we present the AUC that a model has to achieve to meet the high tolerance regime, the loosest criterion in our definition, in different Kaggle competitions.
This is a way to measure the intensity of a competition: The smaller the gap is between the winning AUC and the top 10\% AUC, the more intense the competition is.
Some competitions are highly competitive:
the gap is merely 0.01 on D-EG.

\begin{figure}
\centering
    \includegraphics[width=0.6\columnwidth]{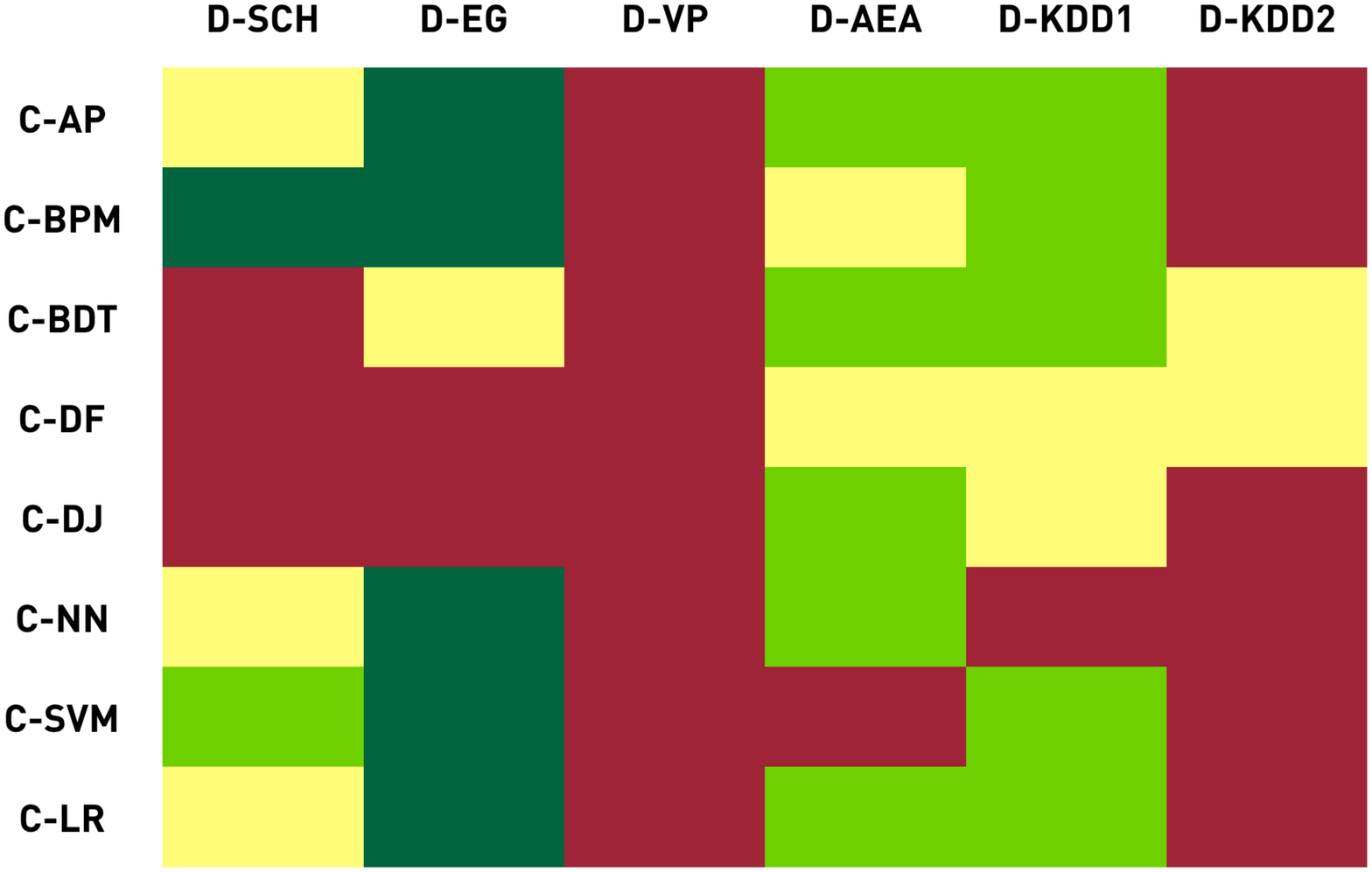}   
\vskip -4ex
\caption{A heat map that represents the capacity of different models on different datasets. Dark green represents low tolerance, light green represents middle tolerance, yellow represents high tolerance and red regions are out of the tolerance regimes we defined.}
\label{fig:heatmap}
\vskip -4ex
\end{figure}

Of course, one can change the thresholds in the above definition and therefore shift the regimes to different regions of the tolerance range $(0, 100]$.
Based on our definition and Figure~\ref{tab:capacity}, {\bf Azure} can meet the low tolerance regime for the datasets D-SCH and D-EG, the middle tolerance regime for the datasets D-AEA and D-KDD1, and the high tolerance regime for the dataset D-KDD2.
In contrast, {\bf Amazon} only meets the high tolerance regime on the dataset D-KDD1 but fails on the others.

To better understand the performance of {\bf Azure} with respect to different quality tolerance regimes, we further present in
Figure~\ref{fig:heatmap} a ``heat map'' that indicates the quality tolerance levels met by different {\bf Azure} models on different datasets (or, in terms of the capacity of models, the heat map represents model capacity across different Kaggle competitions).

The dark green regions correspond to the low tolerance regime, the light green regions correspond to the middle tolerance regime, and the yellow regions correspond to the high tolerance regime.
The other regions are outside the tolerance regimes we defined.
We find that {\bf Azure} actually only meets the low tolerance regime on small datasets, where linear models work well.
{\bf Azure} can meet the middle and high tolerance regimes on large datasets, thanks to the inclusion of nonlinear models.

This is also a summary that covers many observations that we have so far.
In view of the Kaggle competitions (by reading the heat map vertically), some are more challenging than the others.
For example, none of the models can meet even the high tolerance regime on the dataset D-VP, and only C-BDT can meet the high tolerance regime on the dataset D-KDD2.
In view of the models (by reading the heat map horizontally), apparently, there is not a one-size-fits-all solution: No one can dominate the others across the datasets.
Moreover, there is apparently a separation between the ``comfortable zones'' of the models, which we have already stated in Observation~\ref{observation:rule:size}: Linear models are more capable on small datasets, whereas nonlinear models are more capable on large datasets.

\vspace{-1em}
\subsection{Summary and Discussion}

Given the previous analysis, there is an obvious trade-off between the efficiency and effectiveness of machine learning clouds from a user's perspective.
The more alternative models a machine learning cloud provides, the more likely it is that a better model can be found for a particular machine learning task.
However, model selection becomes more challenging and users may spend more time (and money) on finding the most effective model.

Meanwhile, we also find that there is a gap between the best available model on machine learning clouds and the winning code available on Kaggle for certain machine learning tasks.
It is then natural to ask the question of how to narrow the gap
to {\em further improve machine learning clouds.}
Of course, there is no reason to disbelieve that there is a possibility.
For example, one can simply provide more models to increase the chance of finding a better model, though this may make model selection even harder.
It is also not so clear which models should be included, given the trade-off between the capacity of a model and the training time required to tune the model.

We investigated this question from a different viewpoint by looking into the gap itself.
Instead of asking how to make the gap narrower, we ask {\em why} there is a gap.

\begin{figure}
\scriptsize
\centering
\begin{tabular}{c|c|c|r|r}
\hline
Dataset & Best {\bf Azure} & Winning & Quality Gap & Ranking Gap (\%) \\ \hline
D-EG & C-AP & LR &-0.01& -0.3 (No. $4 \rightarrow 2$)\\ 
D-SCH & C-BPM & DWD &-0.01 & 0 \\ 
\hline
\hline
D-KDD1 & C-LR & Ensemble & 0.02 & 2.12 \\ 
D-AEA & C-BDT & Ensemble & 0.01 & 1.6 \\ 
D-KDD2 & C-BDT & Ensemble &0.05 & 6.57 \\ 
D-VP & C-BDT & NA &0.11 & 31.16 \\ 
\hline
\end{tabular}
\caption{Gaps between {\bf Azure} and Kaggle winning code on different datasets (DWD is shorthand for ``distance weighted discrimination'').}
\label{tab:gap}
\vskip -4ex
\end{figure}

\begin{figure*}[t!]
\scriptsize
\centering
\begin{tabular}{lrrrrrrrr|r}
\hline
Dataset &           C-AP &           C-BPM &         C-BDT &          C-DF &          C-DJ &           C-LR &          C-NN &         C-SVM &    BestVsBest \\
\hline

    D-SCH & 0.00 (0.00\%) & 0.00 (0.00\%) & 0.00 (0.00\%) & 0.00 (0.00\%) & 0.00 (0.00\%) & 0.00 (0.00\%) & 0.00 (0.00\%) & 0.00 (0.00\%) & 0.00 (0.00\%) \\
     D-EG & \textcolor{ForestGreen}{0.03 (3.81\%)} & \textcolor{ForestGreen}{0.03 (3.36\%)} & \textcolor{ForestGreen}{0.02 (2.04\%)} & \textcolor{ForestGreen}{0.02 (2.33\%)} & \textcolor{ForestGreen}{0.02 (2.63\%)} & \textcolor{ForestGreen}{0.03 (3.82\%)} & \textcolor{ForestGreen}{0.03 (3.83\%)} & \textcolor{ForestGreen}{0.04 (4.48\%)} & \textcolor{ForestGreen}{0.03 (3.20\%)} \\
     D-VP & \textcolor{ForestGreen}{0.02 (3.89\%)} & \textcolor{ForestGreen}{0.06 (9.75\%)} & \textcolor{ForestGreen}{0.07 (10.71\%)} & \textcolor{ForestGreen}{0.11 (17.64\%)} & \textcolor{ForestGreen}{0.12 (20.66\%)} & \textcolor{ForestGreen}{0.04 (5.41\%)} & \textcolor{ForestGreen}{0.08 (12.25\%)} & \textcolor{ForestGreen}{0.15 (27.24\%)} & \textcolor{ForestGreen}{0.07 (10.71\%)} \\
    D-AEA & \textcolor{ForestGreen}{0.04 (5.14\%)} & \textcolor{ForestGreen}{0.07 (8.42\%)} & \textcolor{ForestGreen}{0.05 (5.73\%)} & \textcolor{ForestGreen}{0.04 (5.26\%)} & \textcolor{ForestGreen}{0.12 (15.31\%)} & \textcolor{ForestGreen}{0.03 (3.88\%)} & \textcolor{ForestGreen}{0.04 (4.46\%)} & \textcolor{ForestGreen}{0.04 (4.74\%)} & \textcolor{ForestGreen}{0.04 (5.03\%)} \\
   D-KDD2 & \textcolor{red}{-0.01 (-1.31\%)} & \textcolor{red}{-0.04 (-8.08\%)} & NA (NA) & \textcolor{ForestGreen}{0.05 (9.07\%)} & \textcolor{ForestGreen}{0.06 (11.80\%)} & \textcolor{red}{-0.02 (-3.06\%)} & NA (NA) & \textcolor{ForestGreen}{0.03 (4.76\%)} & \textcolor{ForestGreen}{0.02 (3.63\%)} \\
   D-KDD1 & \textcolor{ForestGreen}{0.06 (10.60\%)} & \textcolor{ForestGreen}{0.09 (15.35\%)} & NA (NA) & \textcolor{ForestGreen}{0.06 (10.86\%)} & \textcolor{ForestGreen}{0.10 (19.25\%)} & \textcolor{ForestGreen}{0.05 (8.64\%)} & NA (NA) & \textcolor{ForestGreen}{0.10 (17.63\%)} & \textcolor{ForestGreen}{0.05 (8.64\%)} \\
   
\hline
\end{tabular}
\vskip -2ex
\caption{Improvement on private AUC score attributed to feature engineering. Green indicates quality increase
after feature engineering; Red otherwise. All numbers are with hyperparameter tuning. }
\label{tab:feature-engineering:improvement-score}
\vskip -2ex
\end{figure*}

\begin{figure*}[t!]
\scriptsize
\centering
\begin{tabular}{lrrrrrrrr|r}
\hline
Dataset &          C-AP &         C-BPM &        C-BDT &         C-DF &         C-DJ &           C-LR &         C-NN &        C-SVM &   BestVsBest \\
\hline
    D-SCH & 0 (0.00\%) & 0 (0.00\%) & 0 (0.00\%) & 0 (0.00\%) & 0 (0.00\%) & 0 (0.00\%) & 0 (0.00\%) & 0 (0.00\%) & 0 (0.00\%) \\
     D-EG & \textcolor{ForestGreen}{386 (61.76\%)} & \textcolor{ForestGreen}{378 (60.48\%)} & \textcolor{ForestGreen}{320 (51.20\%)} & \textcolor{ForestGreen}{104 (16.64\%)} & \textcolor{ForestGreen}{97 (15.52\%)} & \textcolor{ForestGreen}{386 (61.76\%)} & \textcolor{ForestGreen}{384 (61.44\%)} & \textcolor{ForestGreen}{398 (63.68\%)} & \textcolor{ForestGreen}{372 (59.52\%)} \\
     D-VP & \textcolor{ForestGreen}{59 (4.52\%)} & \textcolor{ForestGreen}{349 (26.72\%)} & \textcolor{ForestGreen}{432 (33.08\%)} & \textcolor{ForestGreen}{627 (48.01\%)} & \textcolor{ForestGreen}{569 (43.57\%)} & \textcolor{ForestGreen}{280 (21.44\%)} & \textcolor{ForestGreen}{527 (40.35\%)} & \textcolor{ForestGreen}{544 (41.65\%)} & \textcolor{ForestGreen}{432 (33.08\%)} \\
    D-AEA & \textcolor{ForestGreen}{759 (44.99\%)} & \textcolor{ForestGreen}{838 (49.67\%)} & \textcolor{ForestGreen}{773 (45.82\%)} & \textcolor{ForestGreen}{742 (43.98\%)} & \textcolor{ForestGreen}{986 (58.45\%)} & \textcolor{ForestGreen}{665 (39.42\%)} & \textcolor{ForestGreen}{726 (43.03\%)} & \textcolor{ForestGreen}{468 (27.74\%)} & \textcolor{ForestGreen}{708 (41.97\%)} \\
   D-KDD2 & \textcolor{red}{-62 (-13.14\%)} & \textcolor{red}{-85 (-18.01\%)} & NA (NA) & \textcolor{ForestGreen}{244 (51.69\%)} & \textcolor{ForestGreen}{301 (63.77\%)} & \textcolor{red}{-121 (-25.64\%)} & NA (NA) & \textcolor{ForestGreen}{157 (33.26\%)} & \textcolor{ForestGreen}{44 (9.32\%)} \\
   D-KDD1 & \textcolor{ForestGreen}{118 (25.00\%)} & \textcolor{ForestGreen}{289 (61.23\%)} & NA (NA) & \textcolor{ForestGreen}{257 (54.45\%)} & \textcolor{ForestGreen}{348 (73.73\%)} & \textcolor{ForestGreen}{65 (13.77\%)} & NA (NA) & \textcolor{ForestGreen}{343 (72.67\%)} & \textcolor{ForestGreen}{65 (13.77\%)} \\

\hline
\end{tabular}
\vskip -2ex
\caption{Improvement on private ranking attributed to feature engineering. Green indicates quality increase after feature engineering; Red otherwise. All numbers are with hyperparameter tuning. }
\label{tab:feature-engineering:improvement-ranking}
\vskip -2ex
\end{figure*}

Figure~\ref{tab:gap} compares the best performing {\bf Azure} model with the winning code from Kaggle.
Again, we separate small datasets (D-EG and D-SCH), where linear models outperform nonlinear models, from large datasets, where nonlinear models are better.
The ``Quality Gap'' column presents the difference in AUC between the winning code and the {\bf Azure} model, and the ``Ranking Gap'' column shows the corresponding movement in rankings on the Kaggle leader board.
For example, on D-EG, the winning code is actually slightly worse than C-AP from {\bf Azure}, with a quality gap of -0.01 and a ranking gap of -0.32\%: The winning code is ranked fourth (i.e., top 0.64\%), whereas C-AP could be ranked second (i.e., top 0.32\%).
The larger the quality gap and ranking gap are, the more potential improvement there is.
One prominent observation from Figure~\ref{tab:gap} is that the winning code on the large datasets leverages ensemble methods (details in Section~\ref{sec:benchmark:dataset-details}), whereas the best nonlinear models from {\bf Azure} (C-BDT, C-DJ, C-DF) more or less leverage ensemble methods as well.
Therefore, it seems that {\bf Azure} is moving in the right direction by supporting more ensemble methods, though it needs to further improve their performance.
{\bf Amazon} may need more work to incorporate ensemble methods (as well as nonlinear models).

\paragraph*{Performance} One angle we intentionally
left out of the picture in this paper is the
performance (speed) of the two machine learning clouds.
Figure~\ref{table:performance} and Figure~\ref{table:performance-average}
contain the training time that Azure needs
for training each model. By modern standards,
these numbers are rather slow --- for the KDD1 dataset,
which contains only 300K training examples and 190
features, training a linear SVM model
takes 205 seconds on average. To fully explore
the hyperparameter space, it is not uncommon
for our experiments to run for hours on 
a single, arguably small, dataset. 
Without knowing the implementation details
of each of these clouds, we intentionally avoid
any discussion and comparison in this paper.
However, these numbers do indicate the possibility
of potential future improvements.

Moreover, Azure currently only allows 
datasets that are smaller than 10GB. For
one Kaggle competition, the winning code
produces a dataset larger than this limitation,
and thus we are not able to benchmark it.
By today's standards, 10GB is a pretty small dataset.
Again, without knowing the implementation details
of Azure cloud, we will leave out concrete comments
about this result. However, we believe it does 
indicate that machine
learning clouds could also be improved on this front.

\paragraph*{Pricing Model} Both machine learning
clouds provide a pricing model based solely
on time --- in our experience, even if we are
willing to pay more money per hour to run our
experiments faster with a beefier machine (or
have more machines to run hyper-parameter tuning
in parallel), it seems that both clouds do not provide
an option. With the diversity of machine learning
tasks and user requirements, we believe
it can also be beneficial to have more flexible 
pricing models in machine learning clouds.

\section{RESULTS ON ALL DATASETS}\label{sec:results:feature-engineering}

So far, our study has been focused on Kaggle competitions
whose winning code is available. We now discuss
the insights we got by analyzing all 18 datasets
in \texttt{mlbench}.

\paragraph*{The Importance of Feature Engineering}

As most of the winning code spends significant effort on feature engineering, there is a clear gap from the typical way that people use machine learning clouds in practice, where feature engineering may not be at the level achieved by the winning code.
Consequently, our previous results for the machine learning clouds may be over-optimistic.
In practice, neither {\bf Azure} or {\bf Amazon}  provides
feature engineering functionality. We assess the
impact of feature engineering and make the case for
a potentially promising research direction of ``declarative
feature engineering on the cloud.''

We consider an extreme case where we do not perform feature engineering, and ask the question:
{\em If we use raw features instead of features constructed by the winning code, how will it impact the performance of machine learning clouds?}
In Figure~\ref{tab:feature-engineering:improvement-score} and~\ref{tab:feature-engineering:improvement-ranking}, we present the improvement in terms of the AUC score and the ranking on the private leader board by using the features from the winning code versus using the raw features
(without feature engineering).
Hyperparameter tuning was turned on for each run.
A negative value here indicates a drop in performance.
Not surprisingly, in most cases using well engineered features helps boost performance significantly, though it is not always the case.
For instance, for C-LR on D-KDD2, using features from the winning code decreases the AUC score by 0.03, and the corresponding ranking on the private leader board drops by 129.
The last columns in Figure~\ref{tab:feature-engineering:improvement-score} and~\ref{tab:feature-engineering:improvement-ranking} further show the improvement by the best model using engineered features versus the best model using raw features.
Even under this best-versus-best comparison, the benefit of feature engineering is significant. 

We also should not be overly pessimistic by the results, though.
After all, in practice it is rare for people to completely give up feature engineering, given the intensive and extensive research on feature selection in the literature.
Consequently, our comparison on using only raw features should be understood as a worst-case study for the performance of machine learning clouds.
Meanwhile, it is interesting to further explore the ``gray areas'' between the two extremes that we have studied in terms of feature engineering: Instead of using either fine-tuned features or just raw features, how will machine learning clouds perform when combined with an
automatic feature learning procedure?
There is apparently a broad spectrum regarding the abundance of feature learning algorithms.
One challenge here is to decide appropriate feature learning procedures for a given learning problem.
Since this is orthogonal (but complementary) to the current work, we leave it as one of the future directions for exploration.
Ideally, this should be integrated into machine learning clouds as part of the declarative service, therefore it might be a promising aspect for machine learning cloud service providers to consider as well.

\begin{figure}[t]
\centering
    \includegraphics[width=0.45\columnwidth]{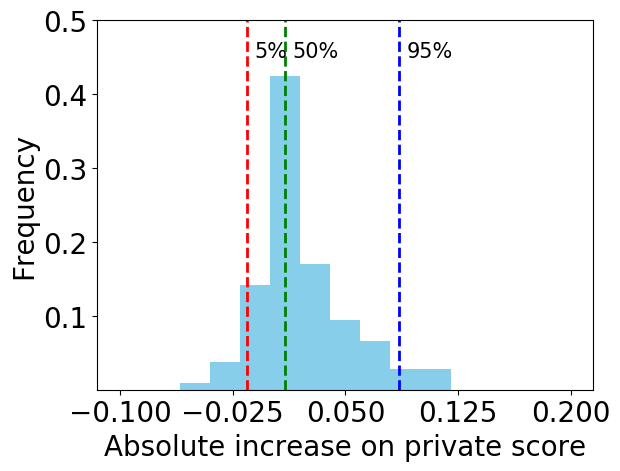}   
    \includegraphics[width=0.45\columnwidth]{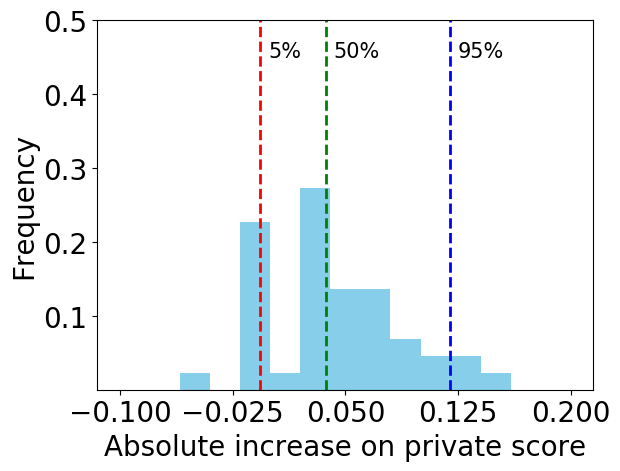}   
    \vskip -2ex
\caption{AUC improvement attributed to (left) hyper-parameter tuning and (right) feature engineering.}
\label{fig:improvement:auc}
\vskip -2ex
\end{figure}

\begin{figure*}
\scriptsize
\centering

\begin{tabular}{l|rrrrrrrr|r}

\hline
Dataset &           C-AP &  C-BPM &        C-BDT &          C-DF &           C-DJ &           C-LR &           C-NN &          C-SVM &     BestVsBest \\
\hline

  D-SCH-r & \textcolor{red}{-0.03 (-3.60\%)} & NA (NA) & \textcolor{ForestGreen}{0.04 (6.79\%)} & \textcolor{ForestGreen}{0.02 (2.39\%)} & \textcolor{ForestGreen}{0.11 (15.87\%)} & \textcolor{ForestGreen}{0.03 (3.92\%)} & \textcolor{red}{-0.01 (-0.81\%)} & \textcolor{ForestGreen}{0.04 (4.47\%)} & \textcolor{ForestGreen}{0.00 (0.22\%)} \\
  D-PIS-r & 0.00 (0.00\%) & NA (NA) & \textcolor{ForestGreen}{0.01 (1.53\%)} & \textcolor{ForestGreen}{0.03 (3.93\%)} & \textcolor{ForestGreen}{0.01 (0.89\%)} & \textcolor{ForestGreen}{0.05 (6.49\%)} & \textcolor{ForestGreen}{0.01 (0.77\%)} & \textcolor{ForestGreen}{0.06 (8.17\%)} & \textcolor{ForestGreen}{0.01 (1.32\%)} \\
   D-EG-r & \textcolor{ForestGreen}{0.00 (0.16\%)} & NA (NA) & \textcolor{ForestGreen}{0.01 (1.12\%)} & \textcolor{ForestGreen}{0.01 (0.81\%)} & \textcolor{ForestGreen}{0.03 (3.25\%)} & \textcolor{red}{-0.00 (-0.04\%)} & \textcolor{ForestGreen}{0.00 (0.16\%)} & \textcolor{ForestGreen}{0.02 (2.65\%)} & \textcolor{ForestGreen}{0.00 (0.16\%)} \\
   D-VP-r & 0.00 (0.00\%) & NA (NA) & \textcolor{ForestGreen}{0.02 (3.66\%)} & \textcolor{ForestGreen}{0.02 (2.84\%)} & \textcolor{red}{-0.00 (-0.32\%)} & \textcolor{red}{-0.02 (-3.40\%)} & \textcolor{ForestGreen}{0.02 (3.07\%)} & \textcolor{red}{-0.02 (-2.94\%)} & \textcolor{red}{-0.01 (-1.56\%)} \\
  D-AEA-r & \textcolor{red}{-0.00 (-0.08\%)} & NA (NA) & \textcolor{ForestGreen}{0.04 (4.20\%)} & \textcolor{ForestGreen}{0.10 (13.74\%)} & \textcolor{ForestGreen}{0.06 (8.68\%)} & \textcolor{ForestGreen}{0.03 (3.22\%)} & \textcolor{ForestGreen}{0.01 (1.34\%)} & \textcolor{ForestGreen}{0.06 (8.09\%)} & \textcolor{ForestGreen}{0.01 (1.12\%)} \\
  D-SCS-r & \textcolor{ForestGreen}{0.00 (0.18\%)} & NA (NA) & \textcolor{ForestGreen}{0.01 (1.00\%)} & \textcolor{ForestGreen}{0.08 (10.65\%)} & \textcolor{ForestGreen}{0.01 (1.57\%)} & \textcolor{ForestGreen}{0.01 (0.80\%)} & \textcolor{ForestGreen}{0.01 (1.45\%)} & \textcolor{red}{-0.01 (-1.56\%)} & \textcolor{ForestGreen}{0.01 (1.00\%)} \\
  D-SMR-r & \textcolor{ForestGreen}{0.00 (0.66\%)} & NA (NA) & NA (NA) & \textcolor{ForestGreen}{0.05 (7.99\%)} & \textcolor{ForestGreen}{0.07 (10.89\%)} & NA (NA) & NA (NA) & NA (NA) & \textcolor{red}{-0.04 (-5.03\%)} \\
  D-GMC-r & \textcolor{ForestGreen}{0.04 (5.77\%)} & NA (NA) & \textcolor{ForestGreen}{0.00 (0.15\%)} & \textcolor{ForestGreen}{0.09 (11.39\%)} & \textcolor{ForestGreen}{0.00 (0.30\%)} & \textcolor{ForestGreen}{0.02 (2.56\%)} & \textcolor{ForestGreen}{0.00 (0.08\%)} & \textcolor{red}{-0.00 (-0.09\%)} & \textcolor{ForestGreen}{0.00 (0.15\%)} \\
  D-HQC-r & 0.00 (0.00\%) & NA (NA) & \textcolor{ForestGreen}{0.00 (0.18\%)} & \textcolor{ForestGreen}{0.03 (3.37\%)} & NA (NA) & 0.00 (0.00\%) & \textcolor{ForestGreen}{0.01 (1.05\%)} & \textcolor{ForestGreen}{0.01 (0.61\%)} & \textcolor{ForestGreen}{0.00 (0.18\%)} \\
  D-KDD-r & \textcolor{ForestGreen}{0.00 (0.43\%)} & NA (NA) & NA (NA) & \textcolor{ForestGreen}{0.05 (10.68\%)} & \textcolor{ForestGreen}{0.03 (6.83\%)} & \textcolor{ForestGreen}{0.01 (1.26\%)} & NA (NA) & \textcolor{red}{-0.01 (-0.95\%)} & \textcolor{ForestGreen}{0.00 (0.83\%)} \\
  D-PBV-r & 0.00 (0.00\%) & NA (NA) & NA (NA) & NA (NA) & NA (NA) & \textcolor{ForestGreen}{0.00 (0.13\%)} & NA (NA) & NA (NA) & 0.00 (0.00\%) \\
  \hline
    D-SCH & \textcolor{red}{-0.03 (-3.60\%)} & NA (NA) & \textcolor{ForestGreen}{0.04 (6.79\%)} & \textcolor{ForestGreen}{0.02 (2.39\%)} & \textcolor{ForestGreen}{0.11 (15.87\%)} & \textcolor{ForestGreen}{0.03 (3.92\%)} & \textcolor{red}{-0.01 (-0.81\%)} & \textcolor{ForestGreen}{0.04 (4.47\%)} & \textcolor{ForestGreen}{0.00 (0.22\%)} \\
     D-EG & \textcolor{ForestGreen}{0.00 (0.42\%)} & NA (NA) & \textcolor{ForestGreen}{0.01 (0.82\%)} & \textcolor{ForestGreen}{0.03 (3.37\%)} & \textcolor{ForestGreen}{0.00 (0.53\%)} & \textcolor{ForestGreen}{0.00 (0.22\%)} & \textcolor{ForestGreen}{0.01 (0.88\%)} & \textcolor{ForestGreen}{0.00 (0.41\%)} & \textcolor{ForestGreen}{0.00 (0.22\%)} \\
     D-VP & \textcolor{red}{-0.05 (-6.70\%)} & NA (NA) & \textcolor{ForestGreen}{0.04 (6.18\%)} & \textcolor{ForestGreen}{0.09 (14.48\%)} & \textcolor{ForestGreen}{0.05 (7.88\%)} & \textcolor{ForestGreen}{0.00 (0.63\%)} & \textcolor{ForestGreen}{0.05 (7.56\%)} & \textcolor{ForestGreen}{0.01 (2.10\%)} & \textcolor{ForestGreen}{0.04 (6.18\%)} \\
    D-AEA & \textcolor{ForestGreen}{0.00 (0.03\%)} & NA (NA) & \textcolor{ForestGreen}{0.01 (1.06\%)} & \textcolor{ForestGreen}{0.07 (8.76\%)} & \textcolor{ForestGreen}{0.01 (1.28\%)} & \textcolor{red}{-0.00 (-0.15\%)} & \textcolor{ForestGreen}{0.04 (4.43\%)} & \textcolor{ForestGreen}{0.00 (0.40\%)} & \textcolor{ForestGreen}{0.01 (0.94\%)} \\
    D-PHY & NA (NA) & NA (NA) & NA (NA) & NA (NA) & NA (NA) & NA (NA) & NA (NA) & NA (NA) & NA (NA) \\
   D-KDD2 & \textcolor{ForestGreen}{0.00 (0.66\%)} & NA (NA) & \textcolor{ForestGreen}{0.00 (0.15\%)} & \textcolor{ForestGreen}{0.07 (13.96\%)} & \textcolor{ForestGreen}{0.09 (17.19\%)} & \textcolor{red}{-0.00 (-0.26\%)} & NA (NA) & \textcolor{red}{-0.00 (-0.75\%)} & \textcolor{ForestGreen}{0.00 (0.15\%)} \\
   D-KDD1 & \textcolor{ForestGreen}{0.03 (5.44\%)} & NA (NA) & \textcolor{red}{-0.02 (-3.07\%)} & \textcolor{ForestGreen}{0.04 (6.27\%)} & \textcolor{ForestGreen}{0.06 (10.96\%)} & \textcolor{ForestGreen}{0.02 (3.87\%)} & \textcolor{ForestGreen}{0.00 (0.34\%)} & \textcolor{ForestGreen}{0.03 (4.84\%)} & \textcolor{red}{-0.01 (-1.56\%)} \\

\hline
\end{tabular}
\vskip -2ex
\caption{Improvement on private AUC score attributed to hyper-parameter tuning. Green indicates quality increase after feature engineering; red otherwise. }
\label{tab:hpt}
\vskip -2ex
\end{figure*}

\paragraph*{The Importance of Hyper-parameter Tuning}

We assess the importance of hyperparamter tuning
in a similar way and the result for
all 18 dataset is in Figure~\ref{tab:hpt}. We see
that hyperparameter tuning has significant impact 
on each individual algorithm, and most of the time,
it improves the quality. One interesting observation
is that, under the best-versus-best comparison (last column),
the impact of hyperparameter tuning drops significantly.
This shows an interesting tradeoff between 
{\em model selection} and {\em hyperparameter
tuning} --- disabling model selection hurts the 
quality more significantly than disabling 
hyperparameter tuning. This opens another interesting
future research question: {\em Given limited computation
budget, how should one balance between model
selection and hyperparameter tuning to maximize 
the final quality?}
Similar to the previous challenge on feature engineering, there are lots of trade-offs.
A naive approach is to perform hyper-parameter tuning for all models and then pick the model with the best performance, exactly as what we have done in our experimental evaluation.
This brute-force approach is perhaps unacceptable in a situation with restrictive resource access.
As another extreme approach, one can first find a model using model selection without any hyper-parameter tuning, and then focus on hyper-parameter tuning for this particular model.
However, there might be little guarantee on the performance of the model selected.
Clearly, there are numerous hybrid strategies in between, where one can first decide on a set of candidate models and then perform hyper-parameter tuning for each candidate.
A declarative machine learning service should hide these details from the user altogether.

We can also compare the impact of hyperparameter tuning
and feature engineering 
by plotting the histograms for the performance improvement
for all individual algorithms.
In Figure~\ref{fig:improvement:auc}, 
the red, green, and blue vertical dashed lines in each figure indicate the 5th, 50th, and 90th percentiles, respectively.
We see that the majority of the improvements falls between 0 and 10\% in terms of AUC score. Comparing
all these three lines ``shifted to the right direction'' 
for the feature engineering plot (i.e., the right histogram), indicating a more significant 
impact attributed to feature engineering than
hyperparameter tuning.





\vspace{-0.5em}
\section{Related Work}\label{sec:relatedwork}

There has been research on benchmarking different machine learning algorithms and comparing their relative quality
on various datasets~\cite{bauer1998empirical,caruana2006empirical,fernandez2014we}. Most of these efforts focus on benchmarking
machine learning algorithms on ``raw datasets'' without much
feature engineering,
a key process for high-quality machine
learning applications~\cite{domingos2012few}. \texttt{mlbench} is different
in the sense that it consists of 
best-effort baselines for feature engineering
and model selection. Another difference between
our study and previous work is that, instead of
benchmarking all existing machine learning models, we 
focus on those provided by existing machine learning clouds
and try to understand whether the current abstraction
is enough to support users of these clouds.

Benchmarking cloud services and more traditional
relational databases have been an active research
topic for decades. Famous benchmarks include the Wisconsin benchmark~\cite{dewittwisconsin} and
TPC benchmarks.\footnote{http://www.tpc.org/information/benchmarks.asp}
There are also benchmarks targeting clouds for different purposes, especially for data processing and management~\cite{cooper2010benchmarking,luo2012cloudrank}.
Our work is motivated by the success and impact of 
these benchmarks, and we hope to establish the first 
benchmark for declarative machine learning on the cloud.

\vspace{-0.5em}
\section{Conclusion}\label{sec:conclusion}

In this paper, we presented an empirical study on the performance of state-of-the-art declarative machine learning clouds.
We conducted our experiments based on \texttt{mlbench}, a dataset we constructed by collecting winning code from Kaggle competitions.
We compared the performance of machine learning clouds with the Kaggle winning code we harvested.
Our results show that there is an obvious gap between top-performing models on the cloud and Kaggle winning code in terms of low quality tolerance regimes they can meet, though machine learning clouds do perform reasonably well when increasing the level of quality tolerance regimes.
Detailed investigation further reveals that lack of adopting ensemble methods is perhaps one reason for the performance gap.
A promising direction for improving the performance of machine learning clouds is therefore to incorporate more well-tuned models that leverage ensemble methods.


\vspace{-0.5em}
{
\scriptsize
\bibliographystyle{abbrv}
\bibliography{vldb_sample}

\begin{thebibliography}{10}

\bibitem{winningcode:schizophrenia}
\url{https://github.com/KKPMW/Kaggle-MLSP-Schizo-3rd}.

\bibitem{winningcode:evergreen}
\url{https://github.com/saffsd/kaggle-stumbleupon2013}.

\bibitem{winningcode:virusprediction}
\url{https://github.com/Cardal/Kaggle_WestNileVirus}.

\bibitem{winningcode:amazon}
\url{https://github.com/owenzhang/Kaggle-AmazonChallenge2013}.

\bibitem{gbm}
\url{https://cran.r-project.org/web/packages/gbm/gbm.pdf}.

\bibitem{random-forest}
\url{https://cran.r-project.org/web/packages/randomForest/randomForest.pdf}.

\bibitem{ert}
\url{https://cran.r-project.org/web/packages/extraTrees/extraTrees.pdf}.

\bibitem{glmnet}
\url{ftp://debian.ustc.edu.cn/CRAN/web/packages/glmnet/glmnet.pdf}.

\bibitem{winningcode:kddcup}
\url{https://www.datarobot.com/blog/datarobot-the-2014-kdd-cup/}.

\bibitem{winningcode:physics}
\url{https://github.com/gramolin/flavours-of-physics}.

\bibitem{amazoncloud}
\url{http://docs.aws.amazon.com/machine-learning/latest/dg/learning-algorithm.html}.

\bibitem{bauer1998empirical}
E.~Bauer and R.~Kohavi.
\newblock An empirical comparison of voting classification algorithms: Bagging,
  boosting, and variants.
\newblock {\em Machine learning}, 1998.

\bibitem{caruana2006empirical}
R.~Caruana et~al.
\newblock An empirical comparison of supervised learning algorithms.
\newblock In {\em ICML}, 2006.

\bibitem{XGBoost}
T.~Chen et~al.
\newblock {XGBoost}: {A} scalable tree boosting system.
\newblock In {\em KDD}, 2016.

\bibitem{cooper2010benchmarking}
B.~F. Cooper, A.~Silberstein, E.~Tam, R.~Ramakrishnan, and R.~Sears.
\newblock Benchmarking cloud serving systems with {YCSB}.
\newblock In {\em SoCC}, 2010.

\bibitem{SVM}
C.~Cortes and V.~Vapnik.
\newblock Support-vector networks.
\newblock {\em Machine Learning}, 1995.

\bibitem{dewittwisconsin}
D.~J. DeWitt.
\newblock The {Wisconsin} benchmark: Past, present, and future.
\newblock In {\em The Benchmark Handbook for Database and Transaction Systems}.
  1993.

\bibitem{domingos2012few}
P.~Domingos.
\newblock A few useful things to know about machine learning.
\newblock {\em CACM}, 2012.

\bibitem{fernandez2014we}
M.~Fern{\'a}ndez-Delgado, E.~Cernadas, S.~Barro, and D.~Amorim.
\newblock Do we need hundreds of classifiers to solve real world classification
  problems.
\newblock {\em JMLR}, 2014.

\bibitem{FreundS97}
Y.~Freund and R.~E. Schapire.
\newblock A decision-theoretic generalization of on-line learning and an
  application to boosting.
\newblock {\em JCSS}, 1997.

\bibitem{hall2009weka}
M.~Hall, E.~Frank, G.~Holmes, B.~Pfahringer, P.~Reutemann, and I.~H. Witten.
\newblock The {WEKA} data mining software: an update.
\newblock In {\em KDD}, 2009.

\bibitem{NN}
S.~Haykin.
\newblock {\em Neural Networks: A Comprehensive Foundation}.
\newblock Prentice Hall PTR, 2nd edition, 1998.

\bibitem{BPM}
R.~Herbrich, T.~Graepel, and C.~Campbell.
\newblock Bayes point machines.
\newblock {\em JMLR}, 2001.

\bibitem{Ho95}
T.~K. Ho.
\newblock Random decision forests.
\newblock In {\em ICDAR}, 1995.

\bibitem{Lui2012}
M.~Lui.
\newblock Feature stacking for sentence classification in evidence-based
  medicine.
\newblock In {\em ALTA}, 2012.

\bibitem{luo2012cloudrank}
C.~Luo et~al.
\newblock Cloudrank-d: benchmarking and ranking cloud computing systems for
  data processing applications.
\newblock {\em Frontiers of Computer Science}, 2012.

\bibitem{DWD}
J.~S. Marron et~al.
\newblock Distance-weighted discrimination.
\newblock {\em JASA}, 2007.

\bibitem{Quinlan86}
J.~R. Quinlan.
\newblock Induction of decision trees.
\newblock {\em Machine Learning}, 1986.

\bibitem{ShottonSKNWC13}
J.~Shotton, T.~Sharp, P.~Kohli, S.~Nowozin, J.~M. Winn, and A.~Criminisi.
\newblock Decision jungles: Compact and rich models for classification.
\newblock In {\em NIPS}, 2013.

\bibitem{zhang2017over}
C.~Zhang, W.~Wu, and T.~Li.
\newblock An overreaction to the broken machine learning abstraction: The
  ease.ml vision.
\newblock In {\em HILDA}, 2017.

\end{thebibliography}
}

\newpage

\end{document}